\documentclass[12pt,preprint]{aastex}
\def \eg        {\hbox{\it e.g.,~}}

\def \vs        {\hbox{\it vs.~}}
\def \etal      {\hbox{\it et~al.}}

\def \arcsec    {{^{\prime\prime}}}

\def \pc        {{\rm\ pc}}
\def \kms       {\hbox{ km s$^{-1}$}}

\def \H0        {{\rm\ H_{0}}}
\def \kmsmpc    {{\rm\ km\ s^{-1}\ Mpc^{-1}}}

\def \ev        {{\rm\ eV}}

\def\gtsim{\raisebox{-.5ex}{$\;\stackrel{>}{\sim}\;$}}
\def \mum       {{\rm\ {\mu}m}}
\def \kel       {{\rm\ K}}

\begin{document}
% \journalid{}{}
% \articleid{}{}
% \slugcomment{submitted to {\it The Astrophysical Journal}}

\shortauthors{Deo, Crenshaw, \& Kraemer}

\shorttitle{{\it Spitzer}/IRS Spectra of Seyfert 1.8s and 1.9s}

\title{Spitzer/IRS Observations of Seyfert 1.8s and 1.9s:\\
  A Comparison with Seyfert 1s and Seyfert 2s}

\author{R. P. Deo\altaffilmark{1}, D. M. Crenshaw\altaffilmark{1},
  S. B. Kraemer\altaffilmark{2}, M. Dietrich\altaffilmark{3},
  M. Elitzur\altaffilmark{4}, H. Teplitz\altaffilmark{5},
  T. J. Turner\altaffilmark{6}}

\altaffiltext{1}{Department of Physics and Astronomy, Georgia State University,
  Atlanta, GA 30303, USA; deo@chara.gsu.edu and crenshaw@chara.gsu.edu}

\altaffiltext{2}{Catholic University of America, and the Exploration of the
  Universe Division, NASA's Goddard Space Flight Center, Code 667,
  Greenbelt, MD 20771, USA; stiskraemer@yancey.gsfc.nasa.gov.}

\altaffiltext{3}{Department of Astronomy, Ohio State University, 4055
  McPherson Laboratory, Columbus, OH 43210, USA;
  dietrich@astronomy.ohio-state.edu}

\altaffiltext{4}{Department of Physics and Astronomy, University of
  Kentucky, Lexington, KY 40506-0055, USA; moshe@pa.uky.edu}

\altaffiltext{5}{Spitzer Science Center, MS 220-6, Caltech, Pasadena, CA
  91125, USA. hit@ipac.caltech.edu}

\altaffiltext{6}{Dept. of Physics, University of Maryland Baltimore County,
  1000 Hilltop Circle, Baltimore, MD 21250, USA
  turner@milkyway.gsfc.nasa.gov ; Code 662, Exploration of the Universe
  Division, NASA/GSFC, Greenbelt, MD 20771, USA}

\begin{abstract}
  We present {\it Spitzer Space Telescope} mid-infrared spectra of 12 Seyfert
  1.8 and 1.9 galaxies over the $5\textrm{--}38 \mum$ region. We compare the
  spectral characteristics of this sample to those of $58$ Seyfert 1 and
  Seyfert 2 galaxies from the {\it Spitzer} archives. An analysis of the
  spectral shapes, the silicate $10 \mum$ feature and the emission line fluxes
  have enabled us to characterize the mid-IR properties of Seyfert 1.8/1.9s.
  We find that the equivalent widths of the $10 \mum$ silicate feature are
  generally weak in all Seyfert galaxies, as previously reported by several
  studies. The few Seyfert galaxies in this sample that show deep $10 \mum$
  silicate absorption features are highly inclined and/or merging galaxies. It
  is likely that these absorption features originate primarily in the dusty
  interstellar medium of the host galaxy rather than in a dusty torus on
  parsec scales close to the central engine. We find that the equivalent width
  of the polycyclic aromatic hydrocarbon (PAH) band at $6.2\mum$ correlates
  strongly with the $20\textrm{--}30\mum$ spectral index. Either of these
  quantities are good indicators of the amount of starburst contribution to
  the mid-IR spectra. The spectra of Seyfert 1.8 and 1.9s are dominated by
  these starburst features, similar to most Seyfert 2s. They show strong PAH
  bands and a strong red continuum toward $30\mum$. The strengths of the
  high-ionization forbidden narrow emission lines [O~IV] $25.89\mum$, [Ne~III]
  $15.56\mum$ and [Ne~V] $14.32\mum$ relative to [Ne~II] $12.81\mum$ are
  weaker in Seyfert 1.8/1.9s and Seyfert 2s as compared to Seyfert 1s. The
  weakness of high-ionization lines in Seyfert 1.8-1.9s is suggestive of
  intrinsically weak active galactic nuclei (AGN) continua, and/or stronger
  star formation activity leading to enhanced [Ne~II]. We discuss the
  implications of these observational results in the context of the Unified
  Model of AGN.
\end{abstract}
\keywords{galaxies: dust, galaxies: active, galaxies: seyfert, spectroscopy:
  infrared}

\section{Introduction}
The study of optical spectra of Seyfert galaxies \citep{1974ApJ...192..581K}
lead to their classification into type 1 and type 2 Seyfert galaxies. Seyfert
1 galaxies have permitted broad ($\sim5000\kms$) and narrow ($\sim500\kms$)
emission lines as well as forbidden narrow emission lines, whereas in Seyfert
2 galaxies the broad emission line component is absent. Study of optical
spectra of type 2 Seyferts using polarized light
\citep{1983ApJ...271L...7M,1985ApJ...297..621A} showed that the broad emission
lines can be detected in some objects. These observations implied that the
structure and physical nature of the central source is similar in all AGN. The
differences between physical properties of type 1 and type 2 Seyfert nuclei
are caused by significant amount of optical obscuration that hides the
broad-line region (BLR) from the line of sight to the observer in type 2
sources. These observations lead to the formation of the Unified Model of AGN
\citep[see reviews by][]{1993ARA&A..31..473A, 1995PASP..107..803U}, where a
dusty structure with a torus-like geometry prevents the direct view of the BLR
when seen at angles close to the equatorial plane of the AGN. The obscuration
within the torus is expected to be composed of dusty molecular clouds (Krolik
and Begelman 1988). However, the exact nature of this obscuration has been a
subject of debate for the last decade \citep[\eg][]{1998ApJS..117...25M}. In
this paper, we report on the mid-infrared {\it Spitzer Space Telescope}
\citep{2004ApJS..154....1W} spectra of an intermediate class of Seyfert
galaxies, Seyfert 1.8s and 1.9s. We compare the mid-IR spectra of this
intermediate class of Seyfert galaxies with Seyfert 1s and Seyfert 2s, and
derive spectral diagnostics to understand the behavior of the dust obscuration
between the subclasses of Seyfert galaxies.

\citet{1981ApJ...249..462O} classified the Seyfert 1 class from 1 to 1.9 with
numerically larger types having weaker broad line components as compared to
the superposed narrow permitted lines. As one goes from a Seyfert 1 to 1.5
class, the superposed permitted narrow emission lines become more
distinguishable from the broad component. The Seyfert 1.8s possess weak broad
wings on H$\alpha$ and H$\beta$, while the 1.9s only show wings on H$\alpha$.
Thus, as we go from a type 1 source to a type 2 source, the waning of broad
emission lines in the intermediate types and the presence of broad lines in
the spectropolarimetry of a few type 2 objects implies that the intermediate
types (1.8/1.9s) may be viewed at intermediate inclination angles to the
central source, possibly through the ``atmosphere'' of the dusty torus.

Further, Seyfert 1.8s and 1.9s show variations in their broad-line Balmer
decrements (H$\alpha/$H$\beta$) over timescales of a few years
\citep{1995ApJ...440..141G}. Goodrich showed that these variations were
consistent with changes in reddening implying a moving dust screen between our
line of sight and the central source. More specifically, the dust screen must
be present between the BLR and the narrow-line region (NLR) as the narrow line
fluxes did not vary over the observed intervals. This is the postulated
location of the dusty torus in AGN. Mid-IR observations of Seyfert 1.8s and
1.9s are thus crucial to constrain the properties of the torus. Further, in
the mid-IR the extinction due to dust should be much lower than in the
optical. Thus we expect the mid-IR spectra of Seyfert 1.8-1.9 to be more like
those of Seyfert 1s than like Seyfert 2s.

Seyfert 1.8s and 1.9s also possess relatively weak high-ionization (\eg
[Fe~VII] $\lambda 6087$) narrow lines as compared to Seyfert 1s, suggesting
that the dust screen may also be obscuring an inner NLR where these lines are
thought to arise \citep{1998ApJ...497L...9M,2000AJ....119.2605N}. The mid-IR
$5\textrm{--}38\mum$ range includes several narrow forbidden emission lines
that arise in the NLR. These lines are useful to test the above hypothesis,
since the extinction of the mid-IR lines is expected to be small.

In the last decade, with the advent of {\it ISO}, several local Seyfert
galaxies have been studied extensively \citep[see,][]{2005SSRv..119..355V}.
Their mid-IR spectral characteristics are well studied
\citep{2000A&A...357..839C,2000A&A...358..481S,2002A&A...393..821S,2004A&A...418..465L}.
Recent studies with the {\it Spitzer Space Telescope}
\citep{2007ApJ...655L..77H, 2007ApJ...654L..49S, 2006AJ....132..401B,
  2005ApJ...633..706W} have focused on obtaining and analyzing the spectra of
local Seyfert galaxies. These observations indicate that most Seyfert 2
galaxies show weak $10 \mum$ silicate absorption, while most Seyfert 1
galaxies show this same feature weakly in emission. Strong silicate absorption
features are clearly seen in local Ultra-Luminous InfraRed Galaxies (ULIRGs)
\citep{2007ApJ...654L..49S}. Most quasars \citep[\eg PG quasars observed by
][]{2005ApJ...625L..75H} tend to show strong $10$ and $18\mum$ silicate
emission bands.

\citet{2006AJ....132..401B} present an IRAS $12\mum$-selected sample of 87
Seyfert 1s and 2s. They present principal component analysis (PCA) of $51$
nuclear spectra extracted from this sample. Using ISO data,
\citet{2000A&A...357..839C} reported a factor of $\sim 8$ difference between
Seyfert 1s and 2s for the continuum at $7\mum$. However,
\citet{2004A&A...418..465L} using ISO data again, showed that the
starburst-corrected continuum luminosities at $6 \mum$ were correlated with
absorption-corrected hard X-ray luminosities, implying no significant
difference between Seyfert 1s and Seyfert 2s. With the {\it Spitzer} InfraRed
Spectrograph \citep[IRS, ][]{2004ApJS..154...18H} observations,
\citet{2006AJ....132..401B} find a factor of $\sim 6$ difference at $6\mum$
between Seyfert 1s and Seyfert 2s. The results of \citet{2006AJ....132..401B}
and \citet{2000A&A...357..839C} suggest that the continuum shortward of $15
\mum$ is suppressed in Seyfert 2s as compared to Seyfert 1s. This is in
agreement with the expected orientation effect in the unification models.
Contemporaneous multiwavelength observations are needed to settle the issue.

Thus, the current body of evidence from IRS spectra suggests that the silicate
strengths are weak and probably are representative of those expected from
clumpy torus models \citep{2002ApJ...570L...9N}. Yet, the behavior of the
$5\textrm{--}18\mum$ spectral slope appears to correlate with optical Seyfert
type: Seyfert 1s have stronger short wavelength mid-IR emission than Seyfert
2s. If, Seyfert 1.8/1.9s are indeed viewed at intermediate angle between
Seyfert 1s and Seyfert 2s, the nature of their mid-IR spectra should reveal
insights into the dust distribution responsible for these trends. In \S 2 we
describe our sample and the data analysis. In \S 3 we present the main
observational results. Finally in \S 4, we summarize the results and discuss
their implications.

\section{Observations and Data Analysis}
We have obtained IRS staring mode observations ({\it Spitzer} general observer
proposal \#3374, PI:  S. Kraemer) of 12 Seyfert 1.8-1.9 galaxies. Dusty
structures are common in the disks of Seyfert host galaxies
\citep{1998ApJS..117...25M,2003ApJ...589..774M} and hence it is important to
minimize the effects of the host galaxy environment on the nuclear mid-IR
spectra. Thus, we chose face-on ($b/a \ge 0.5$) Seyfert 1.8 and 1.9 systems
for this study. We obtained low resolution ($R \sim 64-128$), short low (SL,
$5.2-15.5\mum$) and long low (LL, $15.5 - 38\mum$) spectra. The spectra from
all the modules of the IRS (SL and LL) were obtained in two slit (``nod'')
positions, with varying number of exposure cycles per observation,
commensurate with the mid-IR brightness of the target as estimated from the
$12\mum$ IRAS fluxes. The IRS slits have different slit widths for each
module. The SL2 slit\footnote{Spitzer Observer's Manual,
  http://ssc.spitzer.caltech.edu/documents/SOM/} is
$57\arcsec\times3.6\arcsec$, while SL1 is $57\arcsec\times3.7\arcsec$. The LL2
slit is $168\arcsec\times10.5\arcsec$, while LL1 is
$168\arcsec\times10.7\arcsec$. At a $z$ of $0.01$, $1\arcsec$ corresponds to
about $200\pc$ for $H_{0} = 71 \kmsmpc$. Thus, for a full aperture extraction,
the SL slit will sample about $700\pc$ in the dispersion direction at $z$ of
$0.01$.

The 12 galaxies in our sample are listed in Table~\ref{table1} with their
respective Seyfert type, the ratio of host galaxy minor to major axis ($b/a$)
and their redshifts. The values for the Seyfert type, $b/a$ and redshift were
taken from the NASA Extragalactic Database (NED). Based on the nature of
mid-IR spectra of NGC 7603 and NGC 2622 as compared to mid-IR spectra of
Seyfert 1s \citep{2006AJ....132..401B}, we suggest that these objects are
really Seyfert 1s at the time of observations. These sources are known to have
variable Seyfert type \citep{1976ApJ...210L.117T,1995ApJ...440..141G} and have
transitioned back and forth between Seyfert 1 and Seyfert 1.8 states several
times in past. Mrk 622 was considered to be a Seyfert 1.9
\citep{1995ApJ...440..141G}, but its Seyfert type is listed in NED as Seyfert
2, and its mid-IR continuum is similar to those of Seyfert 2s like Mrk 3.
Thus, this reclassification leads our sample of 12 galaxies to have 9 Seyfert
1.8/1.9s, 2 Seyfert 1s, and 1 Seyfert 2.

To compare our Seyfert 1.8 and 1.9 spectra with the spectra of Seyfert 1s and
Seyfert 2s, we obtained the available {\it Spitzer} archival datasets
originally presented in \citet{2006AJ....132..401B} and
\citet{2005ApJ...633..706W}. The observations by \citet{2006AJ....132..401B}
are IRS spectral maps; hence we have extracted spectra only from the central
slit with the nuclear point source. These comprise a total of 58 Seyfert
galaxies with 19 Seyfert 1s, 4 Seyfert 1.8/1.9s (NGC 4579, NGC 4602, NGC 7314,
and UGC 7064), and 35 Seyfert 2 galaxies. NGC 7603 and UGC 7064 are common
between their sample and our dataset. To this sample, we add observations of
NGC 4151 (Seyfert 1.5) and Mrk 3 (Seyfert 2) from \citet{2005ApJ...633..706W}.
The sample from \citet{2006AJ....132..401B} includes NGC 4151 and we note that
both the spectra match well with each other. Thus, from the archival sample,
we have 19 Seyfert 1s (with NGC 7603 considered to be a Seyfert 1), 36 Seyfert
2s (including Mrk 3) and 3 Seyfert 1.8/1.9s, a total of 58 Seyferts with low
resolution spectra. For this archival sample, we extracted the $b/a$, the
Seyfert type and the redshift from NED.

Combining with our sample with the archival sample, we have 20 Seyfert 1s
(considering NGC 2622 from our sample to be a Seyfert 1), 37 Seyfert 2s (with
Mrk 622 from our sample), 12 Seyfert 1.8/1.9s (3 archival, 9 from our sample),
making a total of 69 Seyferts with low resolution spectra. We also have high
resolution datasets for targets from our sample, but in this paper we restrict
ourselves to only low resolution datasets.

We started with the basic calibrated data (BCD) as processed with the S13.2
pipeline. We used the SMART data analysis system \citep{2004PASP..116..975H}
for the reductions. For low resolution spectra, individual exposures per
``nod'' position were median combined at the image level. The median combined
detector images from one of the orders were then differenced with the ones
from the opposite order (\eg SL2 minus SL1 for first nod position) and
vice-versa to subtract the sky background and to correct rogue pixels. The
spectra were extracted with the ``automatic tapered column point source''
option in SMART with no sky subtraction (since this was done at the image
level). The spectrum for each ``nod'' position was then clipped at the end of
the spectrum where the spectral response function dies off. We cleaned the
data of obviously deviant data points and those data points that were flagged
as bad data from the {\it Spitzer} pipeline. The spectra from each nod
position for a given order (\eg two nod positions for second order SL
spectrum) were then averaged to form the final spectrum for that order. The
final spectra for each order where then scaled with respect to the adjacent
order so that the LL2 spectrum matched the flux level of the LL1 spectrum, SL1
was matched to the scaled LL2 and SL2 was scaled to match the scaled SL1
spectrum. These scale factors typically ranged from $1.02$ to $3.85$ with most
of the scale factors less than $1.2$. The few large scale factors are likely
due to the presence of extended mid-IR emission. The objects most affected
are: NGC 5929, UGC 11680, NGC 1667, MCG-3-34-63, NGC 2639 and MRK 471. The
archival observations from \citet{2006AJ....132..401B} are IRS mapping mode
observations, and hence they have a single ``center nod'' position. The sky
subtraction was done using off-order detector images, in a similar way to the
staring mode observations where we differenced off-order images with same nod
position. The off-order subtraction avoids the problem of differencing nod
positions, where it is possible to subtract the source from itself. The final
spectra were then resampled to a common wavelength grid. An exact match
between spectra extracted from different observing programs for NGC 7603, UGC
7064 and NGC 4151 showed that our spectral extraction process is consistent.

The reduced low resolution IRS spectra of 12 target Seyfert galaxies from our
sample are displayed in Figure~1. We plot the spectra as $F_{\lambda} \ \vs
\lambda$ instead of the customary $F_{\nu}\ \vs \lambda$ as it allows us to
study the spectral features in the $6\textrm{--}15\mum$ range in better
contrast to the $20\textrm{--}30\mum$ range. We do not reproduce the spectra
from \citet{2006AJ....132..401B}, but an overview of the continuum shapes as
seen in $\textrm{F}_{\lambda}$ units is given in Figure~2.

We performed further analysis with generic spectrum analysis programs
available in IDL. We have measured three quantities: the emission line fluxes,
the equivalent widths of the silicate absorption or emission feature and the
PAH features, and the continuum fluxes at 6, 15, 20 and $30\mum$ on the
complete spectra. The continuum fluxes were measured by weighted averaging of
the flux values in a $1\mum$ bin centered at 6, 15, 20 and $30\mum$.
Table~\ref{table2} lists the continuum fluxes measured directly from the
spectra. Table~\ref{table3} lists the equivalent width of the $6.2\mum$ PAH
and the equivalent width of the silicate $10\mum$ feature. Table~\ref{table4}
lists the various narrow emission line fluxes measured for each object from
the low resolution data.

For emission line fluxes, we measured the feature with continuum points
selected on the two sides of the peak of the feature. If the feature was
blended with other features, we have selected the continuum points along the
spectrum such that only the visible contribution of the feature is measured.
We did not attempt deblending of the features which is planned for a future
paper. The flux in the feature was obtained by integrating the area under the
curve above the interpolated local continuum. The features were measured three
times with slightly different continuum points to obtain continuum placement
errors. The 1-$\sigma$ uncertainty of these separate measurements were then
added in quadrature to the uncertainty in the flux measurement as estimated
from 1-$\sigma$ uncertainties on the individual flux points. This value
represents the final uncertainty on the flux measurements. We also measured
the flux of the 6.2 PAH feature.

A related issue here is that the [Ne~II] $12.81\mum$ emission line is blended
with the $12.7\mum$ PAH and the measurement of its flux may be affected due to
this. We have high resolution spectra for 12 (11 from our sample and Mrk 3)
galaxies in our sample. After examining the high resolution spectra, we note
that the [Ne~II] line typically lies between 12.70 to 12.90. We used these
same limits on the low dispersion spectra and measured the line fluxes. The
[Ne~II] line is distinguishable from the $12.7\mum$ PAH feature in the low
resolution spectra. The median value of the absolute difference between the
high and low dispersion measurements for the 12 galaxies is 0.08. However, the
median value of the relative error ($|f_{high}-f_{low}|/f_{high}$) between the
high dispersion and the low dispersion measurement is as high as 47\% for
these 12 galaxies. Assuming that our high-dispersion measurements are correct,
this indicates a significant mismatch between the high resolution and low
resolution line flux measurements. The cause of this mismatch is unclear. The
continuum of both high and low resolution spectra have similar shapes but they
do not necessarily match each other in the scaling; this could be due to
different exposure times, slit widths and the fact that we can not perform sky
subtraction for high resolution data. Also, the region of galaxy sampled by
the high and low dispersion slits are different due to the different
orientations of the slits. A combination of these different observing
conditions may lead to this mismatch. Nevertheless, we find a correlation
coefficient of 0.87 for the high and low resolution flux measurements for the
[Ne~II] line and of 0.91 for the [Ne~III] $15.56\mum$ line. Thus, overall we
estimate that it is possible for [Ne~II] measurements to be affected by
blending with the $12.7\mum$ PAH for low dispersion measurements. This
mismatch between high dispersion and low dispersion fluxes does not appear to
affect the measurements of other lines such as [O~IV] $25.89$, [Ne~III]
$15.56$, and [Ne~V] $14.32\mum$ as they are not blended with a strong PAH
feature.

The equivalent widths were measured using a similar approach to the flux
measurements except that the spectrum was divided by a continuum and the
measurements were done on the continuum-divided spectrum. The continuum was
fitted using continuum flux points at 5.5, 14.5, 20, and $30 \mum$ and a
cubic-spline interpolation. The uncertainty due to the subjective choice of
the continuum contributes to the uncertainty in the measurement of the
equivalent width. We performed a few tests by varying the continuum fit by
small amounts and checking the effects of this on the equivalent widths. Based
on these tests we estimate that the uncertainty due to subjective choice of
the global continuum is $\sim$10\%. When measuring the equivalent width of the
silicate absorption/emission feature, we have not removed the relatively minor
contribution from various emission features in the $8.6\textrm{--}10.8\mum$
range, the most prominent of which are the $9.66\mum\ \textrm{H}_{2}$ emission
line and the [S~IV] $10.51\mum$ line.

\section{Results}
Past studies of the mid-IR spectra of active galaxies from {\it ISO} have lead
to the development of a few spectral diagnostic methods
\citep{2002A&A...393..821S, 2000A&A...359..887L}. Here we present diagnostic
diagrams to characterize the properties of Seyfert 1.8s and 1.9s as compared
to Seyfert 1s and Seyfert 2s using our {\it Spitzer} sample.

\subsection{The Nature of the Mid-IR Spectra of Seyferts}
The Seyfert 1.8/1.9 spectra in Figure~1 show that there are strong variations
in the continuum shapes of Seyferts in the mid-IR. For example, note the
spectrum of Mrk 622, NGC 7603 and UGC 12138. The continuum of UGC 12138 rises
toward shorter wavelengths, but much less steeply than in the case of NGC
7603. In the case of Mrk 622, the continuum diminishes toward shorter
wavelengths. The spectrum of Mrk 334 or of UGC 12138 also show strong PAH
bands, as seen in the spectra of starburst galaxies. In fact, most Seyfert
1.8/1.9 spectra are dominated by these PAH features, in striking similarity to
those of Seyfert 2s with strong starburst contribution
\citep{2006AJ....132..401B}. However, forbidden narrow emission lines such as
[Ne~V] $14.32\mum$ and [O~IV] $25.89\mum$ expected to arise in the NLRs of AGN
indicate presence of an active nucleus in these systems.

The mid-IR spectrum of a Seyfert nucleus is the result of re-emission from the
dust heated by the strong AGN continuum. Thus, these spectra are mainly
dominated by thermal emission due to dust; hot dust ($> 200\kel$) contributes
strongly at shorter wavelengths ($\sim 5\textrm{-}15\mum$), warm dust ($\sim
170\kel$) peaking at $\sim 17\mum$, while cold dust ($\sim 60\kel$)
contributes strongly to the continuum at $\gtsim 30\mum$ . This is strongly
evident in the spectra of Seyfert 1s and Seyfert 2s. Comparing NGC 7603 with
the Seyfert 1 spectra from the archival sample, we note that hot dust
contributes more in Seyfert 1s, while Seyfert 2s like Mrk 3 and Mrk 622 are
dominated by a lack of contribution from hot dust. This lack of contribution
from hot dust can plausibly be attributed to the presence of a torus that is
optically thick at these mid-IR wavelengths. Circumnuclear or extended star
formation contributes independently to both classes as an additional component
in the form of PAH features, low-ionization emission lines, and the cold
thermal component. There are several forbidden low-ionization lines like
[Ne~II] $12.81 \mum$, [S~III] $18.71\mum$, [S~III] $33.48\mum$ and [Si~II]
$34.82\mum$ that trace the starburst component \citep{2002A&A...393..821S}.
Apart from these, there are high-ionization lines like [O~IV] $25.89\mum$,
[Ne~V] $14.32\mum$ and [Ne~V] $24.32\mum$, that are all indicators of strongly
photoionized NLR, revealing the strength of the incident UV/X-ray AGN
radiation field in these systems. Due to the relatively low ionization
potential of $40.96\ev$ for the [Ne~III] $15.56\mum$, it is possible that this
line has a contribution from both AGN and star forming regions. The strengths
of the high-ionization lines like [O IV] $25.89 \mum$ that arise in the NLR
appear to be directly related to the strength of the incident X-ray AGN
continuum (Melendez, M.B. \etal, in preparation).

In Figure~2, we show the variety in continuum shapes of the Seyfert sample
discussed in this paper by plotting a few representative spectra from the
sample. The spectra are scaled to have the same flux at $20\mum$ as that of
Mrk 9. Further, all the spectra are smoothed by a factor of 10 for clarity on
the graph. As can be seen in Figure~2, there are Seyfert galaxies like NGC
7603 or Mrk 9 that show steeping spectra in the $5\textrm{--}15\mum$ range.
The spectra that are similar in shape to NGC 4151 show a break around
$17\mum$. Such a ``break'' or ``hump'' is even more pronounced in spectra like
Mrk 622, where the continuum emission decreases in the $5-15\mum$ range. There
are also spectra like those of NGC 1194 which show strong $10\mum$ absorption
while showing strong continua in the $5-15\mum$ range. Apart from these
variations, there are spectra that show enhanced contribution from star
forming features like PAHs (Mrk 622 and NGC 3079). In such starburst-dominated
sources, there is significant contribution at longer wavelengths
($20\textrm{--}30\mum$) giving rise to the red continuum behavior mentioned by
\citet{2006AJ....132..401B}.

\subsection{Spectral Diagnostics}
In Figures~3--6 we show the trends in spectral characteristics and their
behavior with Seyfert type. In these graphs, Seyfert 1s are represented with
plus symbols, Seyfert 2s are square symbols and Seyfert 1.8/1.9s are filled
triangles. All systems with $b/a \le 0.5$ are circled. Fluxes are measured in
$F_{\lambda}$ units ($\textrm{W}\ \textrm{cm}^{-2}{\mum}^{-1}$) and equivalent
widths are in microns.

Figure~3 is a comparison of $6\textrm{--}15\mum$ and $20\textrm{--}30\mum$
continuum spectral indices. We define the spectral index as $\alpha_{1-2} =
\log(f_{1}(\lambda)/f_{2}(\lambda))/\log(\lambda_{1}/\lambda_{2})$. The
continuum fluxes used in this figure are measured directly from the spectrum
and not from the fitted continuum. A positive $20\textrm{--}30$ spectral index
implies a positive slope on a $log(F_{\lambda})\ \vs log(\lambda)$ plot. In
this graph, the dashed line shows a single power law (see NGC~7603 and Mrk~9
in Figure~2). A more positive $6\textrm{--}15$ and negative $20\textrm{--}30$
(along the arrow labeled BP) results in a broken power law behavior (NGC 4151
and Mrk 622 in Figure~2). Note that there are a number of Seyfert 2s (\eg
Mrk~3) with positive $6\textrm{--}15$ slopes that have more extreme
$6\textrm{--}15$ spectral indices than a Seyfert 1 like NGC~4151. A positive
$20\textrm{--}30$ and a negative $6\textrm{--}15$, along the arrow labeled RC,
leads towards a starburst type spectrum with strong long wavelength continuum
and well defined PAH bands. Most Seyfert 1.8s and 1.9s are mid-way (in
comparison of spectral shapes) between Seyfert 1 types (broken power law,
NGC~4151 and NGC~2622) and starburst type spectra like NGC~3079.

Overall, Figure~3 shows that Seyferts have a wide range of continuum shapes in
the mid-IR. One possible interpretation is that the Seyfert 2 galaxies in the
upper left corner are dominated by components of warm
($T\approx170\textrm{K}$) dust from the nuclear region (\eg the torus or the
inner NLR) that peaks around $17\mum$. These objects have a rather simple
thermal continuum, as shown by Mrk 622 in Figure~1. As one moves down and to
the right in Figure~3, the increasing contribution from starburst-heated dust
(cooler than $\sim 170\ \textrm{K}$) becomes stronger. As one moves from the
upper left to middle left in this diagram, we see a larger contribution from
the hotter dust in the $6\textrm{--}15\mum$ region, possibly due to heating by
the AGN. This is expected for Seyfert 1s, but several Seyfert 2s occupy this
region as well. The majority of Seyfert 1.8/1.9s fall in the region occupied
by Seyfert 2s with significant star formation.

Figure~4 shows the $10\mum$ silicate equivalent width (EW) compared against
the $20\textrm{--}30\mum$ spectral index. Negative EWs imply emission
features. The silicate EWs are {\it small} for most of the Seyferts in the
sample. The mean emission EW in our sample is $-0.16\mum$ with a median of
$-0.08\mum$. The mean absorption EW in our sample is $0.38\mum$ with a median
of $0.26\mum$. \citet{2007ApJ...655L..77H} report silicate strengths
($S_{10\mum} = \ln(F_{\lambda}/F_{cont})$) for a large sample of quasars,
Seyferts and ULIRG galaxies. The quasar and ULIRG silicate strengths bracket
the range of $S_{10\mum}$ for Seyfert galaxies. For quasars, the mean silicate
emission strength reported by \citet{2007ApJ...655L..77H} is $0.20$. For
ULIRGS, it is $-1.56$. For comparison, we measured the apparent silicate
optical depth ($\tau_{\lambda}$) at the peak of the $10\mum$ feature for our
Seyfert sample. The corresponding mean values of silicate strengths
($S_{10\mum}$) are $-0.07$ for Seyfert 1s and $-0.35$ for Seyfert 2s. The
median values are $0.02$ and $-0.18$, respectively. The mean silicate strength
reported by \citet{2007ApJ...655L..77H} for Seyfert 1s and Seyfert 2s, from
their sample, is -0.18 and -0.61. There are several Seyfert 2s and a Seyfert 1
(NGC 1194) in the upper right region of Figure~4. All of these objects are
significantly inclined ($b/a \le 0.5$) and/or are merging with companion
objects. NGC 5256 has a fairly high silicate equivalent width and is a face-on
host galaxy, but is classified in NED as interacting. Thus, their high
equivalent widths are probably a result of absorption in the host galaxy disk
rather than being intrinsic to the AGN. The most important result from this
graph is that Seyferts with inclined or merging host galaxies should be
excluded from studies of the circumnuclear regions and once this is done, the
nuclear regions show only weak silicate emission or absorption features.

Figure~5 shows the $6.2\mum$ PAH equivalent width compared to the
$20\textrm{--}30\mum$ spectral index. The equivalent widths are negative for
emission features and absolute values are plotted on the graph in logarithmic
scale. As can be seen, there is a strong correlation between the
$20\textrm{--}30\mum$ spectral index and the $6.2\mum$ PAH equivalent width.
Increasing $20\textrm{--}30$ spectral index leads to galaxies with increasing
contribution from starbursts. Thus, the PAH contribution is strongly
correlated with the $20\textrm{--}30$ spectral index, which supports our use
of this index as a starburst indicator in Figure~3.

\citet{2007ApJ...655L..77H} find that the silicate strengths correlates with
the ratio of fluxes at $14.5\mum$ and $27.5\mum$ for a large sample of
Seyfert, ULIRG, starburst and QSO sources. The ratio $F(27.5\mum)/F(14.5\mum)$
will trace the continuum contribution of cooler ($T \sim 60\kel$) dust from
the star forming regions, similar to the ratio $F(30\mum)/F(20\mum)$ as seen
in Figure~5. This result in combination with results from Figure~4 and
Figure~5, suggests that inclined host galaxy disks will show the effects of a
geometrically thick, and cool dust distribution. This effect should be taken
into consideration first before invoking other explanations.

Further, \citet{2007ApJ...654L..49S} find that the $6.2\mum$ equivalent width
is anti-correlated with silicate absorption strength for ULIRG and starburst
galaxies. This result hints that deep silicate absorption that is expected to
arise in geometrically and optically thick dust distributions
\citep{2007ApJ...654L..45L} may be connected to highly obscured star forming
regions which tend to show weak/absent PAH features. The absence of strong
silicate absorptions in PAH-dominated starburst spectra, and their presence in
highly inclined and/or merging systems (Figure~4), suggests that the observed
deep silicate absorptions may originate primarily in cool dusty regions on
scales of hundreds of parsecs.

Figure~6 shows the [O IV] $25.89\mum$ to [Ne II] $12.81\mum$ ratio compared to
the [Ne III] $15.56\mum$ to [Ne II] $12.81\mum$ ratio. There is a strong
correlation, with Seyfert 2s and Seyfert 1.8/1.9s showing mostly low [O
IV]/[Ne II] ratios and low [Ne III]/[Ne II] ratios. We see a similar
correlation with [Ne~V]$14.32\mum$/[Ne~II], although there are fewer data
points. Most Seyfert 1s have high [O~IV]/[Ne~II], [Ne~V]/[Ne~II] and
[Ne~III]/[Ne~II]. This correlation for Seyfert 1.8-1.9s and Seyfert 2s
indicates that their NLRs may be subject to similar amounts of ionization,
whereas at least a component of the NLR in Seyfert 1s may experience higher
ionization states, in general. Another alternative that can lead to the
correlation in Figure~6 is that Seyfert 1.8, 1.9 and 2s have more star
formation than Seyfert 1s on a absolute scale, and hence stronger [Ne~II].
There could be two measurement errors that can affect this correlation as
pointed out by the referee: (1) the [Ne~II] line fluxes have a contribution
from adjacent $12.7\mum$ PAH in the low resolution measurements, and (2) the
[O~IV] line is blended with [Fe~II] $25.99\mum$, and hence starburst dominated
systems will have a contribution from the [Fe~II] line as in the case of NGC
3079 \citep{2005ApJ...633..706W}. We discussed the possibility of
contamination of [Ne~II] at the end of the last section, and came to the
conclusion that by carefully measuring the line, without deblending the lines,
our fluxes appear to be consistent with high resolution measurements. We
examined the high and low resolution spectra around the [O~IV] line and
estimated that out of 12 Seyfert galaxies in our sample, only 3 show
comparable contribution from the [Fe~II] line. Without additional high
resolution datasets, we can not rule out that other starburst dominated
systems will show this effect, but it appears to be unlikely if the [O~IV]
line is seen along with [Ne~V] line in the same spectrum. In the case of Mrk
3, we estimate that [Fe~II] contributes less than 10\% of the flux as compared
to [O~IV].

\section{Conclusions}  
We have compared the mid-IR spectral properties of 12 Seyfert 1.8/1.9s with
those of 57 Seyfert 1/1.5s and Seyfert 2s from \citet{2006AJ....132..401B} and
\citet{2005ApJ...633..706W}. The main results are as below.

The Seyfert mid-IR spectra appear to be dominated by differing contributions
of three main thermal components: (1) the hot dust ($T > 200 \kel$) heated by
the nucleus at the inner radius of the dust torus contributing strongly to the
continuum in $5\textrm{--}15\mum$; (2) the warm dust ($T \sim 170\
\textrm{K}$) heated by the nucleus that is expected to be at distances of tens
of parsec from the center; (3) cooler dust ($T \sim 60\kel$) from surrounding
circumnuclear star-forming regions on scales of hundreds of parsec giving rise
to the continuum in the $20\textrm{--}30\mum$ range. This intrinsic form of
the continuum is further modified by: (1) the contribution from PAH emission
of unobscured circum-nuclear star forming regions, and (2) extinction from
cold dust in the host galaxy plane for merging and/or highly inclined host
galaxies. Thus, before we can study extinction due to the dust torus, it is
necessary to exclude AGN with inclined host disks or mergers from such a
study.

The optical classifications of Seyferts do not always carry over to
corresponding spectral shapes in the mid-IR, but there seems to be a general
agreement for Seyfert 1s to show mostly single or broken ``power-law'' type
behavior. Seyfert 2s tend to show a variety of spectral shapes from single and
broken ``power-laws'' to PAH-and-red-continuum dominated spectra. (see
Figure~2), in agreement with trends noted by \cite{2006AJ....132..401B}. The
simplest mid-IR spectra appear to be those of Seyfert 2 galaxies with little
star formation (\eg Mrk 3), which are dominated by a prominent thermal
component that peaks at $\sim17\mum$, leading to the expected $\approx
170\kel$ temperature of this component. We suggest that this component is
present in many Seyfert spectra and becomes evident as the $17\mum$ ``break''.
Seyfert 1s tend to have emission from a hotter component ($T > 200\
\textrm{K}$) in the $6\textrm{--}15\mum$ band and, in many cases, weak
silicate emission, in general agreement with the Unified Model prediction that
we are looking down the hot throat of the dusty torus in type 1 objects.

The equivalent width of the $10\mum$ silicate feature in Seyfert galaxies is
much weaker than expected from uniform density compact torus models (see
Figure~4), confirming previous {\it Spitzer} results by
\citet{2006ApJ...653..127S}, \citet{2007ApJ...654L..49S} and
\citet{2007ApJ...655L..77H}. From the analysis we presented earlier, we note
that the Seyfert galaxies that are highly inclined or interacting with
companions in our sample show strong silicate absorptions.
\citet{2007ApJ...655L..77H} find that the silicate strengths correlate with
the ratio of fluxes at $14.5\mum$ and $27.5\mum$ for a large sample of
Seyfert, ULIRG, starburst and QSO sources. This hints that at least in some
Seyfert galaxies, the strength of the $10\mum$ silicate feature is
significantly affected by contamination from dust in the host galaxy. We note
that this component has been ignored by a number of recent studies.

We find a strong correlation between the $6.2\mum$ PAH equivalent width and
the $20\textrm{--}30\mum$ spectral index (Figure~5). The mid-IR starburst
contribution can thus be characterized using either diagnostic. It appears
that very strong PAHs go together with very red continua toward $30\mum$. On
average, Seyfert 1s tend to show weaker $6.2\mum$ PAH equivalent widths than
Seyfert2s and Seyfert 1.8/1.9s. We note however, that measuring absolute
starburst contribution between Seyfert types will need more work.

In general, the mid-IR spectra of Seyfert 1.8 and 1.9 galaxies are similar in
shape and PAH strengths (Figure~5) with spectra of Seyfert 2 galaxies with
starbursts. This is surprising, since the optical spectra of Seyfert 1.8-1.9s
indicate only a modest amount of reddening of the BLR, typically $E(B-V)
\approx 1$ \citep{1995ApJ...440..141G}. This leads to a relatively small
extinction in the mid-IR, and therefore one would expect that the {\it
  Spitzer} IRS spectra of Seyfert 1.8-1.9s should look more like Seyfert 1s.
We speculate that the enhanced starburst contribution seen in Seyfert 1.8/1.9
spectra may be a combined result of presence of intrinsically stronger star
formation in the {\it Spitzer} aperture and/or weaker intrinsic AGN continuum
in some Seyfert 1.8/1.9s. Further work on this is being done and results will
be reported in an upcoming paper.

In Figure~6 we showed a strong correlation for Seyfert 2s and Seyfert
1.8-1.9s, which tend to show low [O~IV]/[Ne~II] and [Ne~III]/[Ne~II]. There
could be three possible scenarios that could give rise to the observed low
[O~IV]/[Ne~II]: (1) the ionization state of the NLR is low in Seyfert 1.8/1.9s
and Seyfert 2s as compared to Seyfert 1s in the sample, (2) the starburst
component as traced by [Ne~II] and PAH fluxes is stronger in Seyfert 1.8, 1.9
and 2s than in Seyfert 1s, and, (3) these Seyfert 1.8/1.9s have hidden inner
NLRs and are subject to significant obscuration of high-ionization NLR lines.
We can rule out the last scenario for Seyfert 1.8-1.9s. The modest amount of
extinction observed in the optical for Seyfert 1.8-1.9s suggests that the
extinction should be negligible in the mid-IR; thus Seyfert 1.8-1.9s should
have similar [O~IV]/[Ne~II] ratio as Seyfert 1s, which is not observed in
Figure~6. If the starburst contribution is similar between Seyfert 1s and
Seyfert 1.8-1.9s, but the intrinsic AGN continuum is weaker in Seyfert
1.8-1.9s, then that hints at intrinsically low ionization state of the NLRs of
Seyfert 1.8-1.9s. On the other hand, the observed correlation between Seyfert
1.8/1.9s and Seyfert 2s, and the possibility of starburst contribution to
[Ne~III] suggests that similar circum-nuclear starburst contribution between
Seyfert 2s and Seyfert 1.8/1.9s can give rise to this correlation. However,
this does not explain why the Seyfert 1.8/1.9s do not show stronger short
wavelength (hot dust) contribution like Seyfert 1s, in spite of their low
$E(B-V)$ in the optical. An ability to separate absolute contribution of the
active nucleus from the absolute contribution of the starburst is essential
for further progress and we are pursuing this actively.

The AGN type variability and misclassification of Seyfert galaxies may be a
plausible reason for mismatch between the optical Seyfert classification and
the expected mid-IR spectral shapes. Some Seyfert 1.8/1.9 galaxies are known
to transition back-and-forth between a type 1 and a type 2 class object. The
first such detection was for NGC 7603 (one of our targets in this study) by
\citet{1976ApJ...210L.117T}. \citet{1995ApJ...440..141G,1989ApJ...340..190G}
noted that NGC 7603 and NGC 2622 (also one of our targets), transitioned from
a Seyfert 1.8 to a Seyfert 1.5 class possibly as a result of changes in the
intrinsic line of sight reddening. In the mid-IR, they show steepening spectra
towards short wavelengths (see Figure~1) as compared to the rest of our
sample, indicative of Seyfert 1.5 spectra. The mid-IR spectrum of NGC 2622 is
similar in it's continuum shape to the mid-IR spectrum of NGC 4151
\citep[see,][ and Figure~2]{2005ApJ...633..706W}, except that NGC 2622 shows
strong PAH bands. The mid-IR spectrum of NGC 7603 is, however, markedly
different than the rest and has characteristics of a type 1 quasar spectrum
with strong emission bands at $10$ and $18\mum$ and very little cool dust
emission at longer wavelengths. NGC 7603 does however have some starburst
contribution in the form of PAH bands (See Figure~5). Thus, it is important to
compare contemporaneous optical and mid-IR spectra when possible to eliminate
the possibility of type variability in the spectra. For the specific case of
Seyfert 1.8/1.9s discussed in this paper, comparison with contemporaneous
optical spectra and X-ray observations is being done and results will be
reported in a follow-up paper.

\acknowledgments
This work is based in part on observations made with the {\it Spitzer Space
  Telescope} and in part on archival data obtained with the {\it Spitzer Space
  Telescope}, which is operated by the Jet Propulsion Laboratory, California
Institute of Technology under a contract with NASA. Support for this work
was provided by NASA through an award issued by JPL/Caltech. This research
has made use of the NASA/IPAC Extragalactic Database (NED) which is operated
by the Jet Propulsion Laboratory, California Institute of Technology, under
contract with the National Aeronautics and Space Administration. This
research has also made use of NASA's Astrophysics Data System Bibliographic
Services.

{\it Facilities:} \facility{Spitzer}

% 
% 
% References
% \bibliography{ms}

% \clearpage

% 
% Figure captions
% 
\figcaption[fig1a.eps]{{\it Spitzer}/IRS Spectra of Seyfert
  1.8s and 1.9s in $F_{\lambda}\ \vs \lambda$ units (rest frame). PAH
  emission features and narrow emission lines are identified above the
  spectra, and the possible locations of $10\mum$ and $18\mum$ silicate
  emission or absorption are noted below the spectra.}

\figcaption[fig2.eps]{Variations in AGN continuum: Mrk 9 (Seyfert
  1.5, single power law --- SP), NGC 1194 (Seyfert 1, single power law with
  silicate absorption), NGC 4151 (Seyfert 1.5, broken power law --- BP), Mrk
  622 (Seyfert 1.9, broken power law --- BP), NGC 3079 (Seyfert 2, strong PAH,
  strong red continuum --- RC), and NGC 7603 (Seyfert 1.5, unusual quasar like
  spectrum, strong silicate emission feature at $10\mum$). The spectra have
  been normalized to the flux at $20\mum$ and smoothed by a factor of 2. The
  right arrow indicates that as the $20\textrm{--}30\mum$ spectral index
  becomes positive, the mid-IR spectrum is more and more dominated by
  starburst features like PAH bands. The left arrow shows the amount of
  variations in the $6\textrm{--}15\mum$ spectral index, as we go down from
  NGC 7603 to Mrk 622 the contribution of the hottest dust decreases. Optical
  Seyfert classifications do not always agree with the shape of the mid-IR
  spectrum.}

\figcaption[fig3.eps]{$6\textrm{--}15 \mum$ spectral
  index \vs $20\textrm{--}30 \mum$ spectral index. BP --- Broken Power law,
  SP --- Single Power law and RC --- Red Continuum. See Figure~2 for example
  spectra.}

\figcaption[fig4.eps]{Silicate equivalent width
  ($10\mum$) \vs $20\textrm{--}30\mum$ spectral index. The circled symbols
  have host galaxies with $b/a \le 0.5$. NGC 5256 has a fairly high
  equivalent width and is a face-on host galaxy, but is classified as
  interacting (source: NED).}

\figcaption[fig5.eps]{The $6.2\mum$ PAH equivalent
  width \vs the $20\textrm{--}30 \mum$ spectral index. Equivalent widths are
  negative for emission features and absolute values are plotted here on
  logarithmic scale. Going from left to right on the x-axis, starburst
  contribution to the spectrum increases.}

\figcaption[fig6.eps]{[O IV] $25.89$ /[Ne II] $12.81$
  ratio \vs. [Ne III] $15.56$ /[Ne II] $12.81$ ratio. Correlation between
  Seyfert 1.8-1.9s and Seyfert 2s indicates that both subclasses of Seyfert
  galaxies have similar amounts of NLR ionization or they have intrinsically
  stronger starburst contribution as compared to Seyfert 1s.}

\clearpage

% 
% Table 1: Objects
% 
\begin{deluxetable}{lccl}
  \tablewidth{0pt}
  \tablecaption{Sample of Seyfert 1.8 and 1.9 galaxies.\label{table1}}
  \tablehead{\colhead{Galaxy Name} &
    \colhead{$z$} &
    \colhead{$b/a$} &
    \colhead{Seyfert Type}
  }
  \startdata
  Mrk 471   &  0.034  &  0.67  &  Sey 1.8 \\
  Mrk 622   &  0.023  &  0.95  &  Sey $\textrm{2}^{a}$ \\
  Mrk 883   &  0.038  &  1.00  &  Sey 1.8 \\
  NGC 2622  &  0.029  &  0.50  &  Sey $\textrm{1}^{a}$ \\
  Mrk 334   &  0.022  &  0.70  &  Sey 1.8 \\
  UGC 7064  &  0.025  &  1.00  &  Sey 1.8 \\
  Mrk 609   &  0.034  &  0.90  &  Sey 1.8 \\
  NGC 7603  &  0.030  &  0.67  &  Sey $\textrm{1}^{a}$ \\
  UM 146    &  0.017  &  0.77  &  Sey 1.9 \\
  UGC 12138 &  0.025  &  0.88  &  Sey 1.8 \\
  NGC 2639  &  0.011  &  0.61  &  Sey 1.9 \\
  NGC 3786  &  0.009  &  0.59  &  Sey 1.8 \\
  \enddata
  \tablecomments{The redshift of galaxies ($z$), the $b/a$ and the Seyfert
    type are taken from the NASA Extragalactic Database (NED).}
  \tablenotetext{a}{NGC 7603 and NGC 2622 were originally expected to be
    Seyfert 1.8 but have likely transitioned to the Seyfert 1 class, based on
    the appearance of their mid-IR spectra \citep[see also,
    ][]{1976ApJ...210L.117T,1995ApJ...440..141G}. Mrk 622 is classified as a
    Seyfert 2 in NED, and its mid-IR continuum is similar to other Seyfert 2s
    like Mrk 3.}
\end{deluxetable}
% 
% Table 2: Continuum Fluxes
% 
% \include{tab2}
\begin{deluxetable}{llcccc}
  \rotate
  \tabletypesize{\scriptsize}
  \tablewidth{0pt}
  \tablecaption{Continuum flux values (in $F_{\lambda}$ units) from mid-IR spectra.\label{table2}}
  \tablecolumns{6}
  \tablehead{
    \colhead{} &
    \colhead{} & 
    \multicolumn{4}{c}{Continuum flux ($10^{-19}\ \textrm{W}\ \textrm{cm}^{-2}\ {\mu}m^{-1}$) at} \\
    \colhead{Galaxy} &
    \colhead{Seyfert Type$^{a}$} &
    \colhead{$6.00 \mum$} &
    \colhead{$15.00 \mum$} &
    \colhead{$20.00 \mum$} &
    \colhead{$30.00 \mum$}
  }
  \startdata
CGCG381-051  & Sey 2   & 1.66 $\pm$ 0.86   & 1.80 $\pm$ 0.06  & 2.57 $\pm$ 0.05  & 1.92 $\pm$ 0.03  \\
E33-G2       & Sey 2   & 6.36 $\pm$ 0.49   & 3.67 $\pm$ 0.06  & 2.66 $\pm$ 0.10  & 1.23 $\pm$ 0.04  \\
F01475-0740  & Sey 2   & 2.94 $\pm$ 0.59   & 3.43 $\pm$ 0.11  & 3.55 $\pm$ 0.13  & 2.15 $\pm$ 0.05  \\
F04385-0828  & Sey 2   & 15.55 $\pm$ 0.61  & 9.01 $\pm$ 0.29  & 7.92 $\pm$ 0.22  & 6.33 $\pm$ 0.10  \\
F15480-0344  & Sey 2   & 2.95 $\pm$ 0.28   & 3.40 $\pm$ 0.09  & 3.48 $\pm$ 0.08  & 2.23 $\pm$ 0.04  \\
IC4329A      & Sey 1.2 & 35.31 $\pm$ 1.83  & 18.49 $\pm$ 0.46 & 14.30 $\pm$ 0.42 & 6.11 $\pm$ 0.13  \\
Mrk 334      & Sey 1.8 & 6.56 $\pm$ 2.38   & 3.49 $\pm$ 0.13  & 4.31 $\pm$ 0.11  & 4.56 $\pm$ 0.06  \\
Mrk 335      & Sey 1.2 & 8.03 $\pm$ 0.52   & 2.52 $\pm$ 0.08  & 2.00 $\pm$ 0.10  & 1.01 $\pm$ 0.03  \\
Mrk 348      & Sey 2   & 8.06 $\pm$ 0.33   & 5.12 $\pm$ 0.12  & 4.14 $\pm$ 0.18  & 1.90 $\pm$ 0.07  \\
Mrk 471      & Sey 1.8 & 0.96 $\pm$ 0.35   & 0.37 $\pm$ 0.02  & 0.32 $\pm$ 0.02  & 0.30 $\pm$ 0.01  \\
Mrk 6        & Sey 1.5 & 8.69 $\pm$ 0.72   & 4.13 $\pm$ 0.27  & 3.88 $\pm$ 0.12  & 2.06 $\pm$ 0.05  \\
Mrk 609      & Sey 1.8 & 2.97 $\pm$ 1.44   & 1.33 $\pm$ 0.09  & 1.38 $\pm$ 0.03  & 1.45 $\pm$ 0.01  \\
Mrk 622      & Sey 2   & 1.03 $\pm$ 0.40   & 1.51 $\pm$ 0.09  & 1.93 $\pm$ 0.03  & 1.84 $\pm$ 0.02  \\
Mrk 79       & Sey 1.2 & 11.24 $\pm$ 0.64  & 5.20 $\pm$ 0.14  & 4.26 $\pm$ 0.09  & 2.56 $\pm$ 0.06  \\
Mrk 817      & Sey 1.5 & 7.66 $\pm$ 0.46   & 5.11 $\pm$ 0.14  & 5.37 $\pm$ 0.16  & 3.89 $\pm$ 0.07  \\
Mrk 883      & Sey 1.8 & 0.76 $\pm$ 0.27   & 0.64 $\pm$ 0.06  & 0.89 $\pm$ 0.03  & 1.01 $\pm$ 0.01  \\
Mrk 9        & Sey 1.5 & 6.78 $\pm$ 0.51   & 2.68 $\pm$ 0.06  & 2.38 $\pm$ 0.06  & 1.45 $\pm$ 0.04  \\
Mrk 938      & Sey 2   & 10.41 $\pm$ 5.42  & 5.80 $\pm$ 0.16  & 7.68 $\pm$ 0.37  & 14.52 $\pm$ 0.16 \\
M-2-33-34    & Sey 1   & 1.84 $\pm$ 0.45   & 1.61 $\pm$ 0.27  & 1.56 $\pm$ 0.07  & 1.17 $\pm$ 0.01  \\
M-2-40-4     & Sey 2   & 13.95 $\pm$ 0.62  & 6.03 $\pm$ 0.13  & 4.93 $\pm$ 0.10  & 3.58 $\pm$ 0.05  \\
M-2-8-39     & Sey 2   & 2.03 $\pm$ 0.30   & 3.10 $\pm$ 0.07  & 2.39 $\pm$ 0.08  & 0.98 $\pm$ 0.04  \\
M-3-34-63    & Sey 2   & 1.09 $\pm$ 0.60   & 0.28 $\pm$ 0.04  & 0.28 $\pm$ 0.05  & 0.37 $\pm$ 0.01  \\
M-3-58-7     & Sey 2   & 11.74 $\pm$ 0.75  & 4.93 $\pm$ 0.08  & 4.78 $\pm$ 0.08  & 3.33 $\pm$ 0.07  \\
M-5-13-17    & Sey 1.5 & 3.20 $\pm$ 0.44   & 2.79 $\pm$ 0.09  & 2.67 $\pm$ 0.06  & 1.87 $\pm$ 0.03  \\
M-6-30-15    & Sey 1.2 & 12.17 $\pm$ 0.55  & 5.85 $\pm$ 0.10  & 4.84 $\pm$ 0.12  & 2.50 $\pm$ 0.06  \\
NGC 1056     & Sey 2   & 3.75 $\pm$ 2.51   & 1.34 $\pm$ 0.15  & 1.48 $\pm$ 0.11  & 1.79 $\pm$ 0.04  \\
NGC 1125     & Sey 2   & 3.23 $\pm$ 0.85   & 3.12 $\pm$ 0.23  & 3.28 $\pm$ 0.08  & 3.84 $\pm$ 0.03  \\
NGC 1143-4   & Sey 2   & 2.0 $\pm$ 0.43    & $\cdots$         & $\cdots$         & $\cdots$         \\
NGC 1194     & Sey 1   & 13.99 $\pm$ 0.80  & 4.90 $\pm$ 0.16  & 3.20 $\pm$ 0.08  & 2.00 $\pm$ 0.03  \\
NGC 1241     & Sey 2   & 2.15 $\pm$ 0.76   & 1.27 $\pm$ 0.12  & 1.12 $\pm$ 0.12  & 1.12 $\pm$ 0.03  \\
NGC 1320     & Sey 2   & 9.45 $\pm$ 0.39   & 6.36 $\pm$ 0.15  & 5.55 $\pm$ 0.17  & 3.58 $\pm$ 0.07  \\
NGC 1667     & Sey 2   & 3.31 $\pm$ 1.85   & 1.16 $\pm$ 0.11  & 1.08 $\pm$ 0.04  & 1.12 $\pm$ 0.03  \\
NGC 2622     & Sey 1   & 0.84 $\pm$ 0.07   & 0.74 $\pm$ 0.04  & 0.62 $\pm$ 0.02  & 0.34 $\pm$ 0.01  \\
NGC 2639     & Sey 1.9 & 1.44 $\pm$ 0.18   & 0.41 $\pm$ 0.05  & 0.38 $\pm$ 0.02  & 0.44 $\pm$ 0.01  \\
NGC 3079     & Sey 2   & 21.20 $\pm$ 11.51 & 5.54 $\pm$ 0.38  & 3.81 $\pm$ 0.23  & 10.47 $\pm$ 0.32 \\
NGC 3227     & Sey 1.5 & $\cdots$          & 8.61 $\pm$ 0.78  & 8.71 $\pm$ 0.20  & 6.30 $\pm$ 0.12  \\
NGC 3516     & Sey 1.5 & 11.10 $\pm$ 0.66  & 5.12 $\pm$ 0.14  & 4.58 $\pm$ 0.10  & 2.78 $\pm$ 0.05  \\
NGC 3786     & Sey 1.8 & 3.05 $\pm$ 0.68   & 1.23 $\pm$ 0.10  & 1.09 $\pm$ 0.02  & 0.99 $\pm$ 0.01  \\
NGC 3982     & Sey 2   & 1.59 $\pm$ 0.60   & 1.28 $\pm$ 0.11  & 1.43 $\pm$ 0.06  & 1.42 $\pm$ 0.03  \\
NGC 4051     & Sey 1.5 & 14.37 $\pm$ 0.98  & 8.77 $\pm$ 0.22  & 7.39 $\pm$ 0.32  & 4.42 $\pm$ 0.05  \\
NGC 4151     & Sey 1.5 & 47.16 $\pm$ 2.18  & 35.85 $\pm$ 1.88 & 30.08 $\pm$ 0.97 & 12.99 $\pm$ 0.20 \\
NGC 424      & Sey 2   & 35.01 $\pm$ 1.79  & 14.44 $\pm$ 0.25 & 10.36 $\pm$ 0.32 & 4.42 $\pm$ 0.13  \\
NGC 4579     & Sey 1.9 & 3.42 $\pm$ 0.37   & $\cdots$         & $\cdots$         & $\cdots$         \\
NGC 4602     & Sey 1.9 & 1.34 $\pm$ 0.63   & 0.64 $\pm$ 0.04  & 0.74 $\pm$ 0.11  & 0.78 $\pm$ 0.03  \\
NGC 4941     & Sey 2   & $\cdots$          & 1.54 $\pm$ 0.15  & 1.64 $\pm$ 0.08  & 1.24 $\pm$ 0.03  \\
NGC 4968     & Sey 2   & 6.13 $\pm$ 0.94   & 6.26 $\pm$ 0.22  & 5.75 $\pm$ 0.18  & 3.37 $\pm$ 0.05  \\
NGC 5005     & LINER   & 7.91 $\pm$ 1.68   & 2.02 $\pm$ 0.16  & 1.83 $\pm$ 0.12  & 3.40 $\pm$ 0.09  \\
NGC 513      & Sey 2   & 1.97 $\pm$ 0.41   & 1.29 $\pm$ 0.10  & 1.24 $\pm$ 0.09  & 1.05 $\pm$ 0.03  \\
NGC 5256     & Sey 2   & 4.75 $\pm$ 2.79   & 2.35 $\pm$ 0.26  & 2.75 $\pm$ 0.09  & 3.87 $\pm$ 0.01  \\
NGC 526A     & Sey 1.5 & $\cdots$          & 3.60 $\pm$ 0.07  & 2.67 $\pm$ 0.12  & 0.83 $\pm$ 0.02  \\
NGC 5347     & Sey 2   & 4.38 $\pm$ 0.43   & 5.45 $\pm$ 0.13  & 5.04 $\pm$ 0.13  & 2.87 $\pm$ 0.07  \\
NGC 5548     & Sey 1.5 & $\cdots$          & 4.17 $\pm$ 0.09  & 3.68 $\pm$ 0.07  & 1.95 $\pm$ 0.06  \\
NGC 5929     & Sey 2   & 2.73 $\pm$ 0.87   & 0.37 $\pm$ 0.10  & 0.40 $\pm$ 0.05  & 0.48 $\pm$ 0.01  \\
NGC 5953     & Sey 2   & 5.23 $\pm$ 2.94   & 2.77 $\pm$ 0.22  & 2.87 $\pm$ 0.09  & 3.42 $\pm$ 0.07  \\
NGC 7130     & Sey 2   & 6.26 $\pm$ 2.22   & 6.08 $\pm$ 0.27  & 7.69 $\pm$ 0.21  & 9.36 $\pm$ 0.06  \\
NGC 7172     & Sey 2   & $\cdots$          & 3.94 $\pm$ 0.25  & 2.34 $\pm$ 0.07  & 2.86 $\pm$ 0.04  \\
NGC 7314     & Sey 1.9 & $\cdots$          & 2.30 $\pm$ 0.15  & 1.72 $\pm$ 0.07  & 1.33 $\pm$ 0.01  \\
NGC 7469     & Sey 1.2 & 26.83 $\pm$ 8.20  & 19.37 $\pm$ 0.47 & 23.65 $\pm$ 0.17 & 23.83 $\pm$ 0.29 \\
NGC 7496     & Sey 2   & 4.38 $\pm$ 2.27   & 4.26 $\pm$ 0.12  & 6.03 $\pm$ 0.21  & 7.31 $\pm$ 0.07  \\
NGC 7603     & Sey 1   & 14.11 $\pm$ 0.73  & 3.16 $\pm$ 0.08  & 1.86 $\pm$ 0.09  & 0.97 $\pm$ 0.01  \\
NGC 7674     & Sey 2   & 13.19 $\pm$ 0.74  & 9.18 $\pm$ 0.33  & 8.46 $\pm$ 0.18  & 5.78 $\pm$ 0.14  \\
NGC 931      & Sey 1.5 & 14.54 $\pm$ 1.11  & 7.11 $\pm$ 0.20  & 5.70 $\pm$ 0.15  & 3.24 $\pm$ 0.07  \\
TOL 1238-364 & Sey 2   & 6.53 $\pm$ 1.02   & 9.56 $\pm$ 0.29  & 10.98 $\pm$ 0.27 & 7.91 $\pm$ 0.08  \\
UGC 11680    & Sey 2   & 5.18 $\pm$ 0.66   & 2.45 $\pm$ 0.27  & 2.68 $\pm$ 0.42  & 1.60 $\pm$ 0.65  \\
UGC 12138    & Sey 1.8 & 2.49 $\pm$ 0.32   & 1.55 $\pm$ 0.09  & 1.34 $\pm$ 0.04  & 1.00 $\pm$ 0.01  \\
UGC 7064     & Sey 1.8 & 2.57 $\pm$ 0.40   & 1.79 $\pm$ 0.11  & 1.65 $\pm$ 0.06  & 1.29 $\pm$ 0.01  \\
UM 146       & Sey 1.9 & 0.66 $\pm$ 0.05   & 0.44 $\pm$ 0.04  & 0.39 $\pm$ 0.02  & 0.32 $\pm$ 0.01  \\
  \enddata
  \tablecomments{For galaxies NGC 1143-4 and NGC 4579, we could only extract
    the SL spectrum, hence only the measurement of $6\mum$ continuum point is
    given. For galaxy, NGC 4922, we could not extract the complete spectrum,
    hence it is not included in this table.}
  \tablenotetext{a}{Seyfert types are taken from the Nasa Extragalactic
    Database (NED).}
\end{deluxetable}

% 
% Table 3: PAH Fluxes and Silicate EW.
% 
% \include{tab3}
\begin{deluxetable}{lcc}
  \tabletypesize{\scriptsize}
  \tablewidth{0pt}
  \tablecaption{Equivalent width of $6.2 \mum$ PAH and the silicate feature at $9.7 \mum$.\label{table3}}
  \tablecolumns{3}
  \tablehead{
    & 
    \multicolumn{2}{c}{Equivalent Width (in units of $0.1 \mum$)}\\
    \colhead{Galaxy} &
    \colhead{PAH $6.2 \mum$} &
    \colhead{Silicate $9.7 \mum^{a}$}
  }
  \startdata
  CGCG381-051  & -3.73 $\pm$ 0.15  & -2.43 $\pm$ 0.02  \\
  E33-G2       & -0.31 $\pm$ 0.01  & -0.35 $\pm$ 0.01  \\
  F01475-0740  & -1.73 $\pm$ 0.10  & -1.19 $\pm$ 0.86  \\
  F04385-0828  & -0.37 $\pm$ 0.03  & 12.27 $\pm$ 0.04  \\
  F15480-0344  & -0.91 $\pm$ 0.02  & 1.13 $\pm$ 0.15   \\
  IC 4329A     & -0.08 $\pm$ 0.01  & -3.71 $\pm$ 0.14  \\
  Mrk 334      & -3.13 $\pm$ 0.01  & 3.01 $\pm$ 0.04   \\
  Mrk 348      & -0.12 $\pm$ 0.01  & 3.21 $\pm$ 0.05   \\
  Mrk 471      & -3.69 $\pm$ 0.36  & 0.71 $\pm$ 0.43   \\
  Mrk 6        & -0.22 $\pm$ 0.03  & -0.27 $\pm$ 0.04  \\
  Mrk 609      & -4.27 $\pm$ 0.03  & 3.90 $\pm$ 0.03   \\
  Mrk 622      & -3.66 $\pm$ 0.13  & -0.75 $\pm$ 0.10  \\
  Mrk 79       & -0.21 $\pm$ 0.01  & -0.20 $\pm$ 0.07  \\
  Mrk 817      & -0.52 $\pm$ 0.01  & -0.28 $\pm$ 0.03  \\
  Mrk 883      & -5.01 $\pm$ 0.18  & 0.55 $\pm$ 0.04   \\
  Mrk 9        & -0.18 $\pm$ 0.02  & -0.30 $\pm$ 0.06  \\
  Mrk 938      & -5.07 $\pm$ 0.12  & 9.43 $\pm$ 0.53   \\
  M-2-33-34    & -2.83 $\pm$ 0.29  & -0.00 $\pm$ 0.11  \\
  M-2-40-4     & -0.54 $\pm$ 0.02  & 2.54 $\pm$ 0.11   \\
  M-2-8-39     & -0.38 $\pm$ 0.02  & -0.82 $\pm$ 0.19  \\
  M-3-34-63    & -19.48 $\pm$ 1.26 & -15.22 $\pm$ 1.39 \\
  M-3-58-7     & -0.54 $\pm$ 0.02  & -0.21 $\pm$ 0.08  \\
  M-5-13-17    & -1.45 $\pm$ 0.05  & -0.43 $\pm$ 0.20  \\
  M-6-30-15    & -0.09 $\pm$ 0.01  & -0.41 $\pm$ 0.14  \\
  NGC 1056     & -10.01 $\pm$ 0.26 & -4.02 $\pm$ 1.01  \\
  NGC 1125     & -2.95 $\pm$ 0.04  & 12.75 $\pm$ 0.20  \\
  NGC 1143-4   & -2.12 $\pm$ 0.03  & 9.38 $\pm$ 0.45   \\
  NGC 1194     & -0.33 $\pm$ 0.07  & 16.97 $\pm$ 0.06  \\
  NGC 1241     & -5.65 $\pm$ 0.13  & 4.52 $\pm$ 0.23   \\
  NGC 1320     & -0.44 $\pm$ 0.02  & 1.43 $\pm$ 0.04   \\
  NGC 1667     & -8.06 $\pm$ 0.04  & 3.92 $\pm$ 0.09   \\
  NGC 2622     & -1.14 $\pm$ 0.01  & -0.39 $\pm$ 0.06  \\
  NGC 2639     & -0.83 $\pm$ 0.01  & -0.45 $\pm$ 0.01  \\
  NGC 3079     & -5.35 $\pm$ 0.07  & 10.27 $\pm$ 0.45  \\
  NGC 3516     & -0.34 $\pm$ 0.02  & -0.39 $\pm$ 0.08  \\
  NGC 3786     & -1.94 $\pm$ 0.02  & 0.06 $\pm$ 0.13   \\
  NGC 3982     & -0.75 $\pm$ 0.04  & 0.92 $\pm$ 0.22   \\
  NGC 4051     & -0.69 $\pm$ 0.01  & -0.84 $\pm$ 0.15  \\
  NGC 4151     & -0.09 $\pm$ 0.02  & 0.75 $\pm$ 0.12   \\
  NGC 424      & -0.17 $\pm$ 0.01  & 1.39 $\pm$ 0.05   \\
  NGC 4579     & -0.70 $\pm$ 0.02  & -1.29 $\pm$ 0.10  \\
  NGC 4602     & -0.58 $\pm$ 0.03  & 1.50 $\pm$ 0.08   \\
  NGC 4968     & -1.10 $\pm$ 0.04  & 2.65 $\pm$ 0.11   \\
  NGC 5005     & -2.27 $\pm$ 0.06  & 3.57 $\pm$ 0.38   \\
  NGC 513      & -1.69 $\pm$ 0.04  & -0.87 $\pm$ 0.20  \\
  NGC 5256     & -7.24 $\pm$ 0.05  & 9.32 $\pm$ 0.08   \\
  NGC 5347     & -0.74 $\pm$ 0.02  & 1.38 $\pm$ 0.14   \\
  NGC 5548     & $\cdots$          & -1.13 $\pm$ 0.07  \\
  NGC 5929     & -0.22 $\pm$ 0.02  & 0.21 $\pm$ 0.32   \\
  NGC 5953     & -5.11 $\pm$ 0.09  & 1.50 $\pm$ 0.78   \\
  NGC 7130     & -3.93 $\pm$ 0.18  & 4.87 $\pm$ 0.17   \\
  NGC 7469     & -2.80 $\pm$ 0.05  & 2.65 $\pm$ 0.21   \\
  NGC 7496     & -5.15 $\pm$ 0.07  & 1.94 $\pm$ 0.05   \\
  NGC 7603     & -0.74 $\pm$ 0.01  & -1.42 $\pm$ 0.06  \\
  NGC 7674     & -0.44 $\pm$ 0.02  & 2.70 $\pm$ 0.21   \\
  NGC 931      & -0.24 $\pm$ 0.01  & 0.48 $\pm$ 0.04   \\
  TOL 1238-364 & -0.89 $\pm$ 0.06  & 3.49 $\pm$ 0.12   \\
  UGC 11680    & -1.12 $\pm$ 0.03  & -1.39 $\pm$ 0.04  \\
  UGC 12138    & -1.08 $\pm$ 0.02  & 1.68 $\pm$ 0.03   \\
  UGC 7064     & -1.10 $\pm$ 0.11  & 1.68 $\pm$ 0.01   \\
  UM 146       & -1.34 $\pm$ 0.11  & 0.13 $\pm$ 0.02   \\
  \enddata
  \tablecomments{Emission features have negative equivalent widths. The errors
    quoted here are 1-$\sigma$ measurement errors as propagated from error
    vectors returned by SMART during spectral reductions. There will be an
    additional $10\%$ error due to subjective placement of the continuum.}

  \tablenotetext{a}{The silicate equivalent width includes a contribution from
    the $\textrm{H}_{2}$ emission feature at $9.66\mum$ and in some cases from
    the [S~IV] $10.51\mum$ emission line. No attempt has been made to subtract
    these contributions.}
\end{deluxetable}

% 
% Table 4: Line Ratios
% 
\pagestyle{empty}
\begin{deluxetable}{lccccccccccc}
  \rotate
  \tablewidth{1.4\textwidth}
  \tabletypesize{\scriptsize}
  \tablecaption{Integrated fluxes for prominent narrow emission lines and the $6.2\mum$ PAH feature.\label{table4}}
  \tablecolumns{12}
  \tablehead{
    \colhead{} &
    \multicolumn{11}{c}{Integrated line fluxes ($10^{-19} W cm^{-2}$) at} \\
    \colhead{Galaxy} &
    \colhead{PAH} &
    \colhead{[Ar II]} &
    \colhead{[Ar III]} &
    \colhead{[S IV]} &
    \colhead{[Ne II]} &
    \colhead{[Ne V]} &
    \colhead{[Ne III]} &
    \colhead{[S III]} &
    \colhead{[Ne V]} &
    \colhead{[O IV]} &
    \colhead{[S III]} \\
    \colhead{} &
    \colhead{$6.2 \mum$} &
    \colhead{$6.99 \mum$} &
    \colhead{$8.99 \mum$} &
    \colhead{$10.51 \mum$} &
    \colhead{$12.81 \mum$} &
    \colhead{$14.32 \mum$} &
    \colhead{$15.56 \mum$} &
    \colhead{$18.71 \mum$} &
    \colhead{$24.32 \mum$} &
    \colhead{$25.89 \mum$} &
    \colhead{$33.48 \mum$}
  }
  \startdata
 CGCG381-051 & 0.52 $\pm$  0.04 & $\cdots$         & 0.05 $\pm$  0.01 & $\cdots$         & 0.17 $\pm$  0.01 & $\cdots$         & 0.03 $\pm$  0.01 & 0.09 $\pm$  0.01 & $\cdots$         & 0.01 $\pm$  0.01 & 0.13 $\pm$  0.02 \\
 E33-G2      & 0.14 $\pm$  0.04 & $\cdots$         & 0.07 $\pm$  0.01 & 0.05 $\pm$ 0.01  & $\cdots$         & 0.08 $\pm$ 0.01  & 0.12 $\pm$ 0.01  & 0.07 $\pm$ 0.01  & 0.06 $\pm$ 0.01  & 0.15 $\pm$ 0.01  & 0.03 $\pm$ 0.01  \\
 F01475-0740 & 0.44 $\pm$  0.04 & 0.04 $\pm$  0.01 & 0.05 $\pm$  0.02 & 0.03 $\pm$  0.01 & 0.16 $\pm$  0.01 & $\cdots$         & 0.12 $\pm$  0.01 & 0.09 $\pm$  0.01 & 0.03 $\pm$  0.01 & 0.07 $\pm$  0.01 & 0.06 $\pm$  0.02 \\
 F04385-0828 & 0.25 $\pm$  0.02 & $\cdots$         & 0.09 $\pm$  0.03 & $\cdots$         & 0.24 $\pm$  0.02 & $\cdots$         & 0.18 $\pm$  0.02 & $\cdots$         & 0.10 $\pm$  0.01 & 0.12 $\pm$  0.02 & $\cdots$         \\
 F15480-0344 & $\cdots$         & 0.08 $\pm$  0.01 & $\cdots$         & 0.07 $\pm$  0.01 & 0.07 $\pm$  0.01 & 0.10 $\pm$  0.01 & 0.15 $\pm$  0.01 & 0.09 $\pm$  0.01 & 0.09 $\pm$  0.01 & 0.34 $\pm$  0.01 & $\cdots$         \\
 IC4329A     & 0.27 $\pm$  0.04 & 0.07 $\pm$  0.02 & 0.08 $\pm$  0.01 & 0.46 $\pm$  0.02 & 0.26 $\pm$  0.01 & 0.43 $\pm$  0.02 & 0.70 $\pm$  0.03 & 0.19 $\pm$  0.02 & 0.37 $\pm$  0.02 & 1.08 $\pm$  0.04 & 0.18 $\pm$  0.02 \\
 Mrk 3       & $\cdots$         & 0.30 $\pm$  0.07 & $\cdots$         & 0.51 $\pm$  0.08 & 0.86 $\pm$  0.12 & 1.09 $\pm$  0.12 & 2.07 $\pm$  0.29 & 0.59 $\pm$  0.11 & 0.59 $\pm$  0.07 & 2.10 $\pm$  0.26 & 0.82 $\pm$  0.20 \\
 Mrk 334     & 1.55 $\pm$  0.06 & 0.17 $\pm$  0.02 & $\cdots$         & 0.11 $\pm$  0.02 & 0.30 $\pm$  0.08 & 0.13 $\pm$  0.02 & 0.26 $\pm$  0.03 & 0.23 $\pm$  0.05 & 0.11 $\pm$  0.02 & 0.15 $\pm$  0.03 & 0.90 $\pm$  0.08 \\
 Mrk 335     & 0.11 $\pm$  0.03 & 0.03 $\pm$  0.02 & $\cdots$         & 0.07 $\pm$  0.01 & 0.05 $\pm$  0.02 & $\cdots$         & $\cdots$         & $\cdots$         & $\cdots$         & 0.09 $\pm$  0.01 & $\cdots$         \\
 Mrk 348     & 0.16 $\pm$  0.01 & $\cdots$         & $\cdots$         & 0.06 $\pm$  0.01 & 0.13 $\pm$  0.01 & $\cdots$         & 0.25 $\pm$  0.01 & 0.04 $\pm$  0.01 & 0.02 $\pm$  0.01 & 0.24 $\pm$  0.01 & 0.05 $\pm$  0.01 \\
 Mrk 471     & 0.25 $\pm$  0.02 & $\cdots$         & $\cdots$         & 0.03 $\pm$  0.01 & 0.03 $\pm$  0.01 & 0.02 $\pm$  0.01 & 0.03 $\pm$  0.01 & 0.06 $\pm$  0.01 & 0.01 $\pm$  0.01 & 0.04 $\pm$  0.01 & 0.07 $\pm$  0.01 \\
 Mrk 6       & 0.20 $\pm$  0.02 & $\cdots$         & 0.08 $\pm$  0.02 & 0.22 $\pm$  0.01 & 0.24 $\pm$  0.01 & 0.12 $\pm$  0.01 & 0.47 $\pm$  0.01 & 0.21 $\pm$  0.01 & $\cdots$         & 0.50 $\pm$  0.01 & 0.07 $\pm$  0.02 \\
 Mrk 609     & 0.93 $\pm$  0.04 & $\cdots$         & $\cdots$         & 0.05 $\pm$  0.01 & 0.12 $\pm$  0.02 & 0.09 $\pm$  0.01 & 0.09 $\pm$  0.02 & 0.13 $\pm$  0.02 & $\cdots$         & 0.08 $\pm$  0.01 & 0.17 $\pm$  0.04 \\
 Mrk 622     & 0.27 $\pm$  0.02 & $\cdots$         & $\cdots$         & $\cdots$         & 0.06 $\pm$  0.02 & $\cdots$         & 0.08 $\pm$  0.02 & 0.05 $\pm$  0.01 & $\cdots$         & 0.10 $\pm$  0.01 & $\cdots$         \\
 Mrk 79      & 0.22 $\pm$  0.01 & $\cdots$         & 0.07 $\pm$  0.01 & 0.11 $\pm$  0.01 & 0.10 $\pm$  0.01 & $\cdots$         & 0.20 $\pm$  0.01 & 0.04 $\pm$  0.01 & 0.05 $\pm$  0.01 & 0.51 $\pm$  0.01 & $\cdots$         \\
 Mrk 817     & 0.36 $\pm$  0.01 & $\cdots$         & $\cdots$         & 0.09 $\pm$  0.01 & 0.05 $\pm$  0.01 & 0.08 $\pm$  0.01 & 0.04 $\pm$  0.01 & 0.03 $\pm$  0.01 & $\cdots$         & 0.06 $\pm$  0.01 & $\cdots$         \\
 Mrk 883     & 0.22 $\pm$  0.02 & $\cdots$         & 0.01 $\pm$  0.01 & 0.03 $\pm$  0.01 & 0.11 $\pm$  0.01 & $\cdots$         & 0.08 $\pm$  0.01 & 0.08 $\pm$  0.02 & $\cdots$         & 0.10 $\pm$  0.01 & 0.07 $\pm$  0.02 \\
 Mrk 9       & $\cdots$         & 0.04 $\pm$  0.01 & 0.05 $\pm$  0.01 & 0.07 $\pm$  0.01 & 0.02 $\pm$  0.01 & $\cdots$         & 0.01 $\pm$  0.01 & $\cdots$         & 0.05 $\pm$  0.01 & 0.07 $\pm$  0.01 & 0.03 $\pm$  0.01 \\
 Mrk 938     & 3.53 $\pm$  0.05 & 0.18 $\pm$  0.03 & $\cdots$         & $\cdots$         & 0.64 $\pm$  0.02 & $\cdots$         & 0.20 $\pm$  0.02 & $\cdots$         & $\cdots$         & $\cdots$         & 0.37 $\pm$  0.03 \\
 M-2-33-34   & 0.42 $\pm$  0.04 & 0.08 $\pm$  0.02 & 0.04 $\pm$  0.02 & 0.26 $\pm$  0.02 & 0.14 $\pm$  0.01 & 0.15 $\pm$  0.01 & 0.40 $\pm$  0.01 & 0.06 $\pm$  0.02 & 0.06 $\pm$  0.01 & 0.60 $\pm$  0.01 & 0.15 $\pm$  0.03 \\
 M-2-40-4    & 0.73 $\pm$  0.04 & $\cdots$         & 0.02 $\pm$  0.02 & 0.04 $\pm$  0.01 & 0.14 $\pm$  0.01 & 0.12 $\pm$  0.01 & 0.18 $\pm$  0.02 & 0.04 $\pm$  0.01 & $\cdots$         & 0.18 $\pm$  0.01 & $\cdots$         \\
 M-2-8-39    & $\cdots$         & $\cdots$         & $\cdots$         & 0.05 $\pm$  0.01 & 0.07 $\pm$  0.01 & 0.10 $\pm$  0.01 & 0.10 $\pm$  0.01 & 0.04 $\pm$  0.01 & $\cdots$         & 0.15 $\pm$  0.01 & $\cdots$         \\
 M-3-34-63   & 0.38 $\pm$  0.10 & $\cdots$         & $\cdots$         & 0.02 $\pm$  0.02 & 0.19 $\pm$  0.05 & $\cdots$         & $\cdots$         & $\cdots$         & $\cdots$         & $\cdots$         & $\cdots$         \\
 M-3-58-7    & 0.55 $\pm$  0.04 & $\cdots$         & 0.02 $\pm$  0.01 & 0.03 $\pm$  0.02 & 0.05 $\pm$  0.04 & $\cdots$         & 0.10 $\pm$  0.01 & 0.06 $\pm$  0.01 & 0.10 $\pm$  0.01 & 0.12 $\pm$  0.01 & $\cdots$         \\
 M-5-13-17   & 0.36 $\pm$  0.18 & $\cdots$         & $\cdots$         & 0.08 $\pm$  0.01 & 0.11 $\pm$  0.01 & $\cdots$         & 0.13 $\pm$  0.01 & 0.01 $\pm$  0.01 & $\cdots$         & 0.12 $\pm$  0.01 & 0.15 $\pm$  0.01 \\
 M-6-30-15   & 0.05 $\pm$  0.02 & 0.04 $\pm$  0.02 & 0.08 $\pm$  0.02 & 0.09 $\pm$  0.01 & 0.04 $\pm$  0.01 & $\cdots$         & 0.01 $\pm$  0.01 & $\cdots$         & $\cdots$         & 0.20 $\pm$  0.02 & 0.11 $\pm$  0.02 \\
 NGC 1056    & 2.06 $\pm$  0.05 & $\cdots$         & 0.04 $\pm$  0.01 & 0.04 $\pm$  0.01 & 0.23 $\pm$  0.01 & $\cdots$         & 0.25 $\pm$  0.01 & 0.28 $\pm$  0.03 & $\cdots$         & $\cdots$         & 0.14 $\pm$  0.01 \\
 NGC 1125    & 0.70 $\pm$  0.06 & $\cdots$         & 0.04 $\pm$  0.02 & 0.12 $\pm$  0.05 & 0.27 $\pm$  0.08 & $\cdots$         & 0.31 $\pm$  0.01 & 0.18 $\pm$  0.01 & 0.03 $\pm$  0.01 & 0.29 $\pm$  0.01 & 0.07 $\pm$  0.06 \\
 NGC 1143-4  & 0.30 $\pm$  0.03 & $\cdots$         & 0.03 $\pm$  0.03 & 0.07 $\pm$  0.01 & 0.15 $\pm$  0.01 & $\cdots$         & $\cdots$         & $\cdots$         & $\cdots$         & $\cdots$         & $\cdots$         \\
 NGC 1194    & 0.18 $\pm$  0.03 & $\cdots$         & 0.03 $\pm$  0.03 & 0.19 $\pm$  0.03 & 0.05 $\pm$  0.02 & 0.12 $\pm$  0.02 & 0.09 $\pm$  0.01 & $\cdots$         & 0.06 $\pm$  0.01 & 0.14 $\pm$  0.01 & $\cdots$         \\
 NGC 1241    & 0.71 $\pm$  0.05 & $\cdots$         & 0.03 $\pm$  0.01 & 0.13 $\pm$  0.02 & 0.09 $\pm$  0.01 & $\cdots$         & $\cdots$         & $\cdots$         & $\cdots$         & 0.02 $\pm$  0.01 & 0.18 $\pm$  0.04 \\
 NGC 1320    & 0.48 $\pm$  0.01 & $\cdots$         & 0.04 $\pm$  0.01 & 0.13 $\pm$  0.01 & 0.09 $\pm$  0.01 & 0.08 $\pm$  0.01 & 0.09 $\pm$  0.01 & 0.13 $\pm$  0.01 & 0.10 $\pm$  0.01 & 0.32 $\pm$  0.01 & $\cdots$         \\
 NGC 1667    & 1.05 $\pm$  0.01 & 0.21 $\pm$  0.01 & 0.13 $\pm$  0.01 & $\cdots$         & 0.26 $\pm$  0.01 & $\cdots$         & 0.19 $\pm$  0.01 & 0.20 $\pm$  0.01 & $\cdots$         & 0.12 $\pm$  0.01 & 0.10 $\pm$  0.01 \\
 NGC 2622    & 0.03 $\pm$  0.01 & $\cdots$         & 0.02 $\pm$  0.01 & 0.03 $\pm$  0.01 & 0.06 $\pm$  0.01 & $\cdots$         & $\cdots$         & 0.02 $\pm$  0.01 & 0.02 $\pm$  0.01 & 0.08 $\pm$  0.02 & 0.02 $\pm$  0.02 \\
 NGC 2639    & 0.10 $\pm$  0.02 & $\cdots$         & $\cdots$         & $\cdots$         & 0.11 $\pm$  0.02 & $\cdots$         & 0.06 $\pm$  0.01 & 0.05 $\pm$  0.01 & $\cdots$         & 0.03 $\pm$  0.01 & 0.11 $\pm$  0.02 \\
 NGC 3079     & 7.79 $\pm$  0.12 & 0.13 $\pm$  0.03 & $\cdots$         & 0.16 $\pm$  0.05 & 1.48 $\pm$  0.09 & $\cdots$         & 0.28 $\pm$  0.01 & 0.38 $\pm$  0.02 & $\cdots$         & 0.36 $\pm$  0.02 & 0.65 $\pm$  0.07 \\
 NGC 3227     & $\cdots$         & $\cdots$         & $\cdots$         & $\cdots$         & $\cdots$         & $\cdots$         & 0.91 $\pm$ 0.03  & 0.41 $\pm$ 0.02  & 0.18 $\pm$ 0.02  & 0.73 $\pm$ 0.01  & 0.19 $\pm$ 0.02  \\
 NGC 3516     & 0.26 $\pm$  0.01 & 0.06 $\pm$  0.01 & 0.03 $\pm$  0.01 & 0.17 $\pm$  0.01 & 0.08 $\pm$  0.01 & 0.11 $\pm$  0.01 & 0.26 $\pm$  0.01 & 0.11 $\pm$  0.01 & 0.09 $\pm$  0.01 & 0.38 $\pm$  0.01 & 0.24 $\pm$  0.01 \\
 NGC 3786     & 0.65 $\pm$  0.03 & $\cdots$         & 0.04 $\pm$  0.01 & 0.08 $\pm$  0.01 & 0.11 $\pm$  0.01 & $\cdots$         & 0.12 $\pm$  0.03 & 0.11 $\pm$  0.01 & 0.03 $\pm$  0.02 & 0.19 $\pm$  0.02 & 0.15 $\pm$  0.04 \\
 NGC 3982     & 0.21 $\pm$  0.01 & 0.06 $\pm$  0.01 & 0.02 $\pm$  0.01 & 0.06 $\pm$  0.01 & 0.16 $\pm$  0.01 & $\cdots$         & 0.05 $\pm$  0.01 & 0.04 $\pm$  0.01 & $\cdots$         & 0.02 $\pm$  0.01 & 0.12 $\pm$  0.01 \\
 NGC 4051     & 0.90 $\pm$  0.03 & 0.09 $\pm$  0.02 & 0.01 $\pm$  0.01 & 0.03 $\pm$  0.01 & 0.17 $\pm$  0.02 & 0.18 $\pm$  0.02 & 0.23 $\pm$  0.02 & 0.11 $\pm$  0.02 & 0.17 $\pm$  0.02 & 0.32 $\pm$  0.01 & 0.17 $\pm$  0.02 \\
 NGC 4151     & 0.69 $\pm$  0.04 & 0.44 $\pm$  0.03 & 0.37 $\pm$  0.03 & 0.86 $\pm$  0.02 & 1.35 $\pm$  0.03 & 1.08 $\pm$  0.06 & 2.50 $\pm$  0.11 & 1.14 $\pm$  0.05 & 0.52 $\pm$  0.12 & 2.45 $\pm$  0.07 & 0.46 $\pm$  0.04 \\
 NGC 424      & 1.16 $\pm$  0.05 & 0.12 $\pm$  0.01 & 0.08 $\pm$  0.01 & 0.26 $\pm$  0.01 & 0.15 $\pm$  0.01 & 0.11 $\pm$  0.01 & 0.21 $\pm$  0.01 & 0.06 $\pm$  0.01 & 0.09 $\pm$  0.01 & 0.18 $\pm$  0.01 & 0.32 $\pm$  0.01 \\
 NGC 4579     & 0.26 $\pm$  0.03 & 0.05 $\pm$  0.02 & 0.01 $\pm$  0.01 & $\cdots$         & 0.11 $\pm$  0.01 & $\cdots$         & $\cdots$         & $\cdots$         & $\cdots$         & $\cdots$         & $\cdots$         \\
 NGC 4602     & 0.14 $\pm$  0.02 & 0.11 $\pm$  0.04 & 0.07 $\pm$  0.02 & $\cdots$         & 0.16 $\pm$  0.02 & 0.04 $\pm$  0.01 & $\cdots$         & 0.09 $\pm$  0.01 & $\cdots$         & 0.05 $\pm$  0.01 & 0.23 $\pm$  0.02 \\
 NGC 4941     & $\cdots$         & $\cdots$         & $\cdots$         & $\cdots$         & $\cdots$         & 0.09 $\pm$ 0.01  & 0.20 $\pm$ 0.01  & 0.15 $\pm$ 0.01  & 0.07 $\pm$ 0.01  & 0.19 $\pm$ 0.04  & 0.15 $\pm$ 0.10  \\
 NGC 4968     & 0.65 $\pm$  0.03 & $\cdots$         & 0.04 $\pm$  0.02 & 0.05 $\pm$  0.01 & 0.24 $\pm$  0.01 & 0.17 $\pm$  0.02 & 0.24 $\pm$  0.05 & 0.19 $\pm$  0.02 & 0.10 $\pm$  0.02 & 0.30 $\pm$  0.01 & 0.32 $\pm$  0.02 \\
 NGC 5005     & 1.63 $\pm$  0.06 & $\cdots$         & 0.04 $\pm$  0.01 & $\cdots$         & 0.60 $\pm$  0.01 & $\cdots$         & 0.16 $\pm$  0.01 & 0.08 $\pm$  0.01 & $\cdots$         & 0.18 $\pm$  0.01 & 0.20 $\pm$  0.02 \\
 NGC 513      & 0.26 $\pm$  0.03 & $\cdots$         & 0.07 $\pm$  0.01 & 0.01 $\pm$  0.01 & 0.11 $\pm$  0.01 & $\cdots$         & 0.08 $\pm$  0.01 & 0.09 $\pm$  0.01 & $\cdots$         & 0.09 $\pm$  0.01 & 0.18 $\pm$  0.02 \\
 NGC 5256     & 2.02 $\pm$  0.08 & 0.13 $\pm$  0.02 & $\cdots$         & 0.09 $\pm$  0.01 & 0.76 $\pm$  0.02 & $\cdots$         & 0.37 $\pm$  0.01 & 0.26 $\pm$  0.01 & 0.17 $\pm$  0.02 & 0.61 $\pm$  0.01 & 0.29 $\pm$  0.03 \\
 NGC 526A     & $\cdots$         & $\cdots$         & $\cdots$         & $\cdots$         & $\cdots$         & 0.05 $\pm$ 0.01  & 0.13 $\pm$ 0.01  & 0.10 $\pm$ 0.01  & $\cdots$         & 0.21 $\pm$ 0.01  & 0.07 $\pm$ 0.02  \\
 NGC 5347     & 0.21 $\pm$  0.03 & $\cdots$         & $\cdots$         & 0.03 $\pm$  0.01 & 0.03 $\pm$  0.01 & $\cdots$         & $\cdots$         & 0.03 $\pm$  0.01 & $\cdots$         & 0.04 $\pm$  0.01 & 0.07 $\pm$  0.01 \\
 NGC 5548     & $\cdots$         & $\cdots$         & $\cdots$         & 0.02 $\pm$  0.02 & 0.04 $\pm$  0.02 & 0.05 $\pm$  0.02 & 0.14 $\pm$  0.01 & 0.04 $\pm$  0.01 & 0.04 $\pm$  0.01 & 0.15 $\pm$  0.01 & 0.04 $\pm$  0.01 \\
 NGC 5929     & 0.55 $\pm$  0.17 & $\cdots$         & 0.07 $\pm$  0.03 & 0.05 $\pm$  0.02 & 0.21 $\pm$  0.02 & 0.04 $\pm$  0.01 & 0.09 $\pm$  0.01 & $\cdots$         & $\cdots$         & 0.07 $\pm$  0.01 & 0.11 $\pm$  0.01 \\
 NGC 5953     & 1.88 $\pm$  0.05 & 0.06 $\pm$  0.02 & $\cdots$         & $\cdots$         & 1.05 $\pm$  0.02 & $\cdots$         & 0.21 $\pm$  0.01 & 0.61 $\pm$  0.02 & $\cdots$         & 0.22 $\pm$  0.01 & 0.69 $\pm$  0.02 \\
 NGC 7130     & 1.77 $\pm$  0.08 & 0.22 $\pm$  0.03 & 0.04 $\pm$  0.01 & 0.07 $\pm$  0.01 & 0.71 $\pm$  0.05 & 0.11 $\pm$  0.02 & 0.37 $\pm$  0.02 & 0.37 $\pm$  0.10 & $\cdots$         & 0.43 $\pm$  0.03 & 0.64 $\pm$  0.03 \\
 NGC 7172     & $\cdots$         & $\cdots$         & $\cdots$         & $\cdots$         & $\cdots$         & $\cdots$         & 0.19 $\pm$ 0.01  & 0.14 $\pm$ 0.02  & 0.10 $\pm$ 0.01  & 0.50 $\pm$ 0.02  & 0.41 $\pm$ 0.02  \\
 NGC 7314     & $\cdots$         & $\cdots$         & $\cdots$         & $\cdots$         & $\cdots$         & 0.23 $\pm$ 0.01  & 0.24 $\pm$ 0.01  & 0.15 $\pm$ 0.01  & 0.19 $\pm$ 0.01  & 0.53 $\pm$ 0.01  & 0.19 $\pm$ 0.02  \\
 NGC 7469     & 5.36 $\pm$  0.15 & 0.78 $\pm$  0.04 & $\cdots$         & 0.15 $\pm$  0.02 & 1.68 $\pm$  0.04 & $\cdots$         & 0.45 $\pm$  0.02 & 1.24 $\pm$  0.03 & $\cdots$         & 1.03 $\pm$  0.16 & 2.05 $\pm$  0.08 \\
 NGC 7496     & 1.41 $\pm$  0.03 & $\cdots$         & 0.02 $\pm$  0.01 & 0.01 $\pm$  0.01 & 0.34 $\pm$  0.01 & 0.05 $\pm$  0.01 & 0.03 $\pm$  0.01 & 0.34 $\pm$  0.02 & $\cdots$         & $\cdots$         & 0.32 $\pm$  0.02 \\
 NGC 7603     & 1.10 $\pm$  0.04 & $\cdots$         & 0.03 $\pm$  0.01 & 0.07 $\pm$  0.01 & 0.11 $\pm$  0.01 & $\cdots$         & 0.07 $\pm$  0.01 & 0.13 $\pm$  0.01 & 0.06 $\pm$  0.01 & 0.07 $\pm$  0.01 & 0.10 $\pm$  0.03 \\
 NGC 7674     & 0.50 $\pm$  0.04 & $\cdots$         & $\cdots$         & 0.22 $\pm$  0.01 & 0.18 $\pm$  0.01 & 0.31 $\pm$  0.02 & 0.46 $\pm$  0.02 & 0.22 $\pm$  0.02 & 0.24 $\pm$  0.02 & 0.46 $\pm$  0.02 & 0.07 $\pm$  0.01 \\
 NGC 931      & 0.44 $\pm$  0.10 & $\cdots$         & $\cdots$         & 0.21 $\pm$  0.02 & 0.08 $\pm$  0.01 & 0.18 $\pm$  0.02 & 0.22 $\pm$  0.02 & 0.09 $\pm$  0.02 & 0.26 $\pm$  0.02 & 0.43 $\pm$  0.02 & 0.10 $\pm$  0.01 \\
 TOL 1238-364 & 0.63 $\pm$  0.09 & 0.05 $\pm$  0.01 & 0.04 $\pm$  0.01 & 0.05 $\pm$  0.01 & 0.39 $\pm$  0.02 & 0.09 $\pm$  0.03 & 0.34 $\pm$  0.03 & 0.23 $\pm$  0.02 & $\cdots$         & 0.12 $\pm$  0.01 & 0.53 $\pm$  0.03 \\
 UGC 11680    & 0.53 $\pm$  0.09 & $\cdots$         & $\cdots$         & $\cdots$         & 0.20 $\pm$  0.04 & $\cdots$         & $\cdots$         & $\cdots$         & $\cdots$         & $\cdots$         & 0.15 $\pm$  0.04 \\
 UGC 12138    & 0.26 $\pm$  0.02 & 0.03 $\pm$  0.01 & $\cdots$         & 0.06 $\pm$  0.01 & 0.03 $\pm$  0.01 & 0.06 $\pm$  0.01 & 0.08 $\pm$  0.01 & 0.04 $\pm$  0.01 & $\cdots$         & 0.11 $\pm$  0.02 & 0.04 $\pm$  0.01 \\
 UGC 7064     & 0.09 $\pm$  0.02 & $\cdots$         & 0.03 $\pm$  0.01 & $\cdots$         & 0.07 $\pm$  0.01 & $\cdots$         & 0.10 $\pm$  0.01 & 0.07 $\pm$  0.01 & 0.07 $\pm$  0.01 & 0.17 $\pm$  0.01 & 0.06 $\pm$  0.01 \\
 UM 146       & $\cdots$         & 0.02 $\pm$  0.01 & $\cdots$         & 0.02 $\pm$  0.01 & 0.03 $\pm$  0.01 & 0.02 $\pm$  0.01 & 0.04 $\pm$  0.01 & 0.02 $\pm$  0.01 & $\cdots$         & 0.02 $\pm$  0.01 & 0.05 $\pm$  0.01 \\
  \enddata
\end{deluxetable}

\clearpage
\pagestyle{plain}

% 
% Figures
% 
% Figure 1: Spectra
\setcounter{figure}{0}
\begin{figure}[!ht]
  \vspace{-2.0cm}
  \begin{center}
    \includegraphics[width=0.45\textwidth,angle=90]{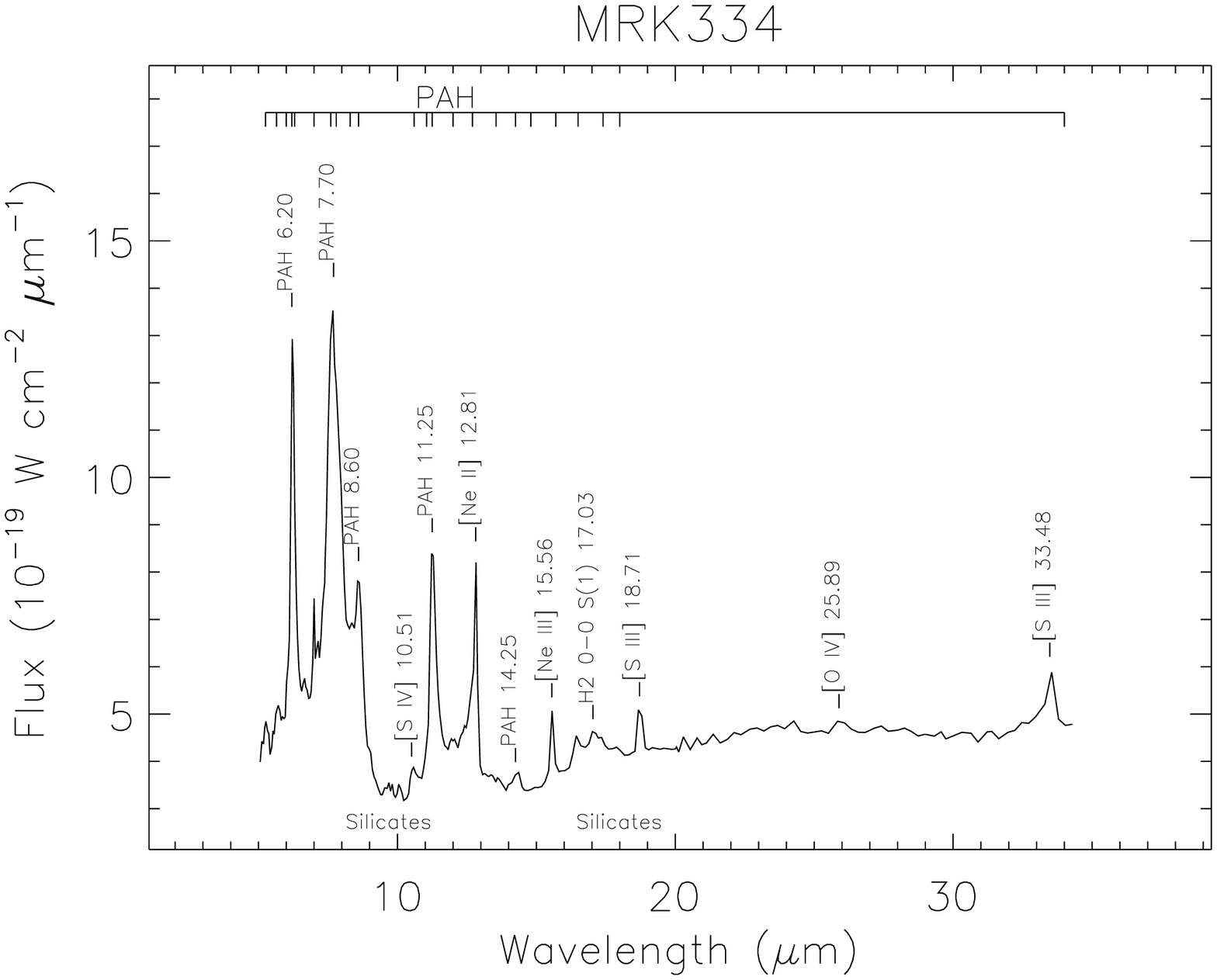}\\
    \includegraphics[width=0.45\textwidth,angle=90]{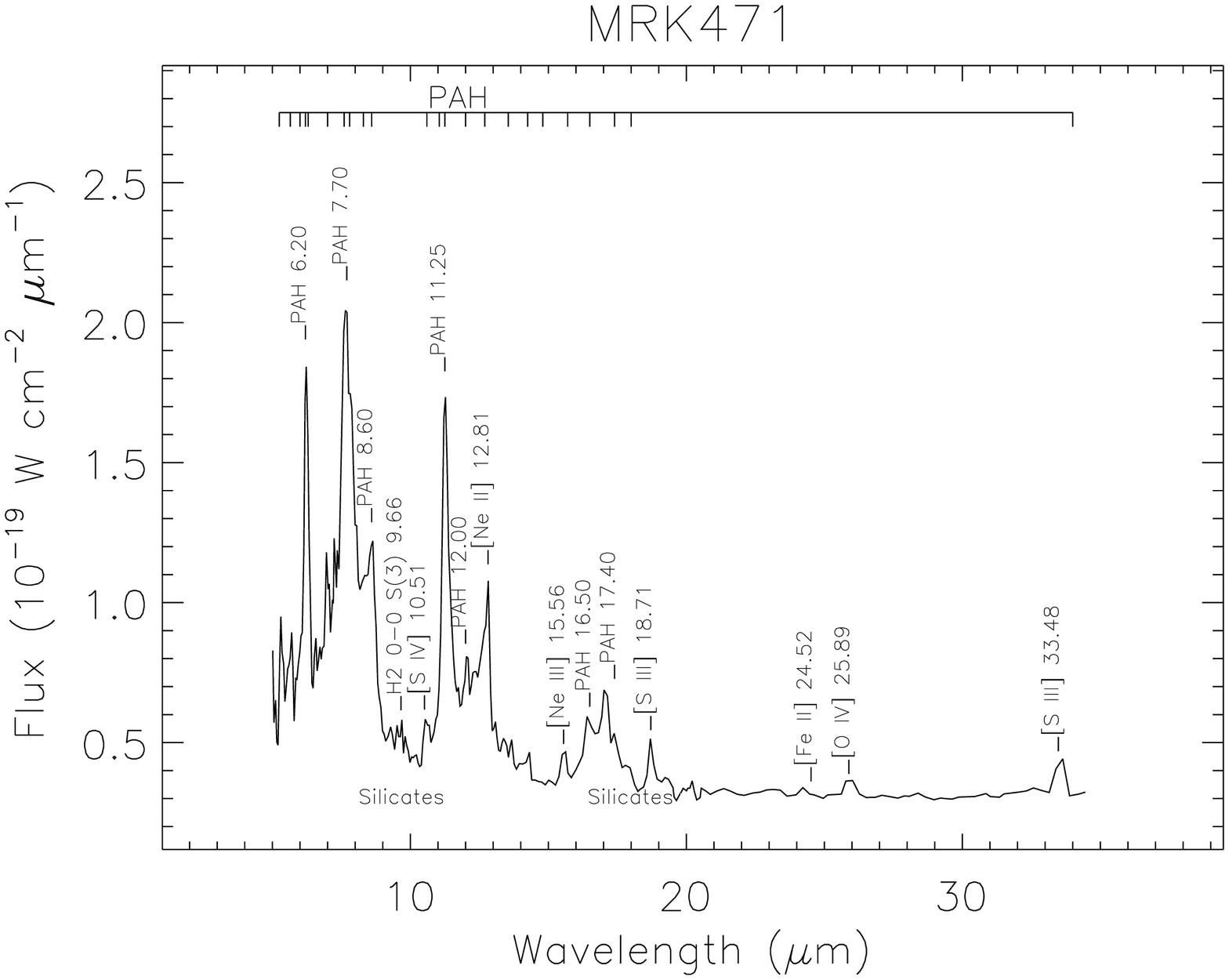}\\
    \includegraphics[width=0.45\textwidth,angle=90]{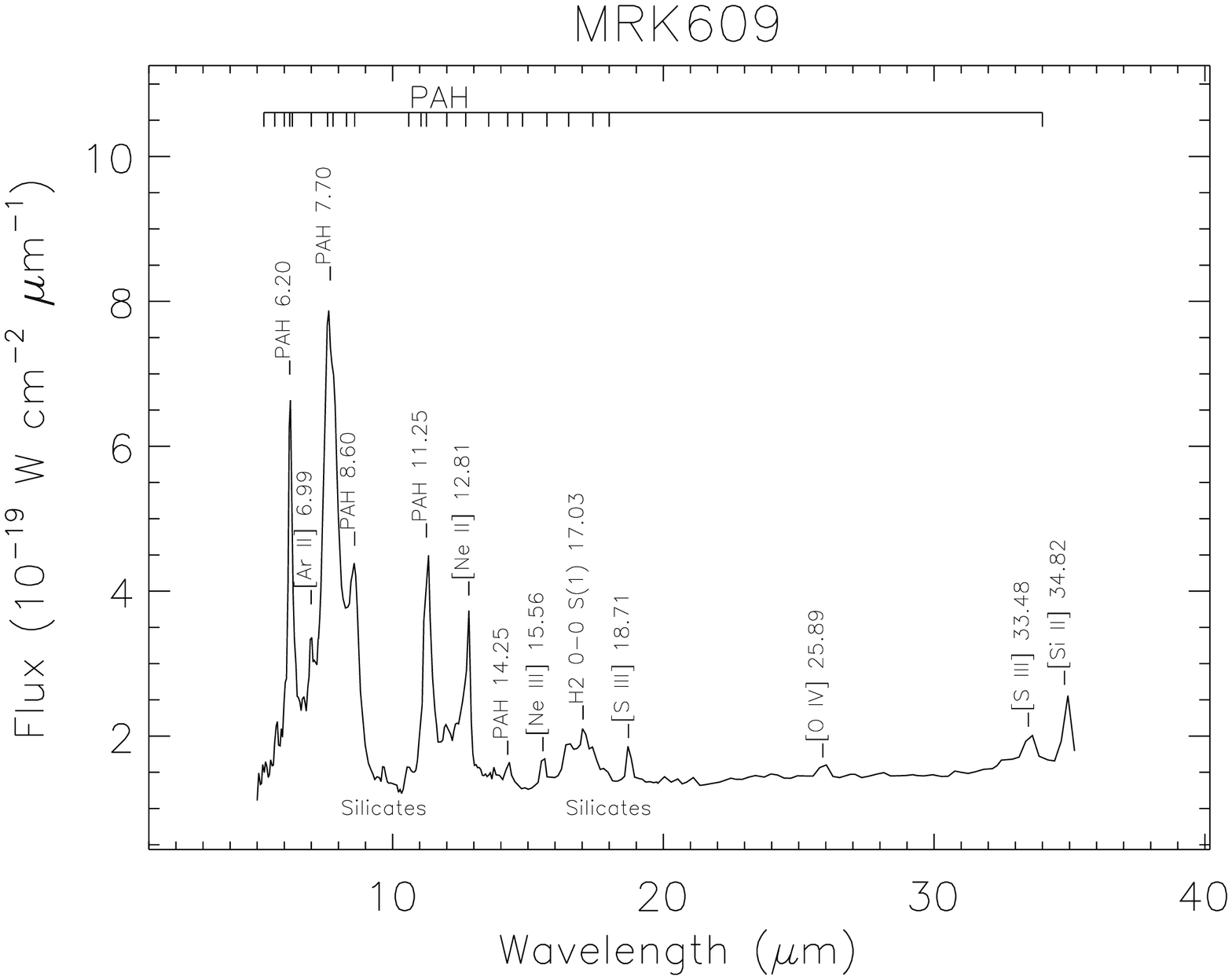}\\
    \caption{Spectra of Seyferts from our new observations}
  \end{center}
\end{figure}
\clearpage

\setcounter{figure}{0}
\begin{figure}[!ht]
  \vspace{-2.0cm}
  \begin{center}
    \includegraphics[width=0.45\textwidth,angle=90]{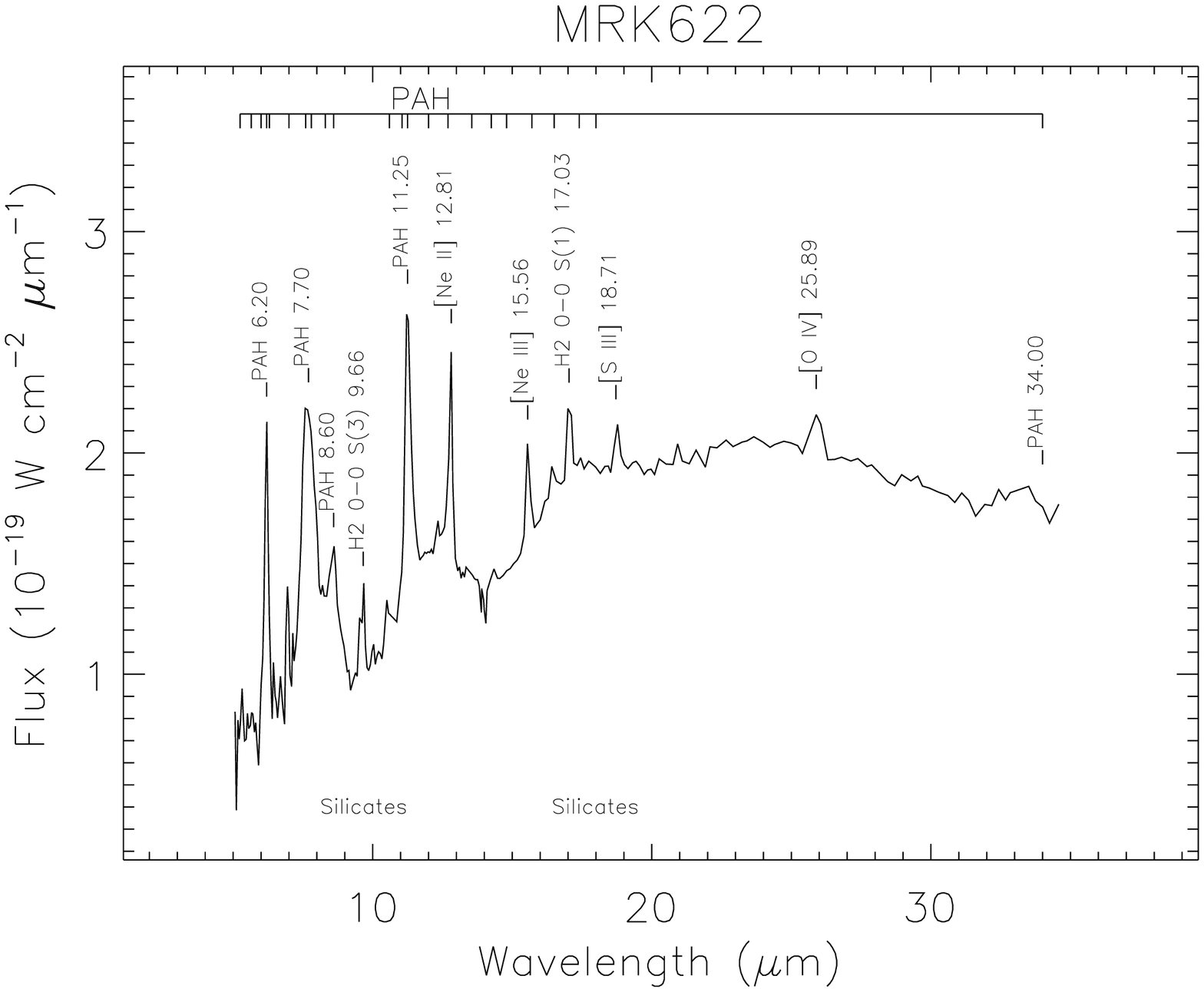}\\
    \includegraphics[width=0.45\textwidth,angle=90]{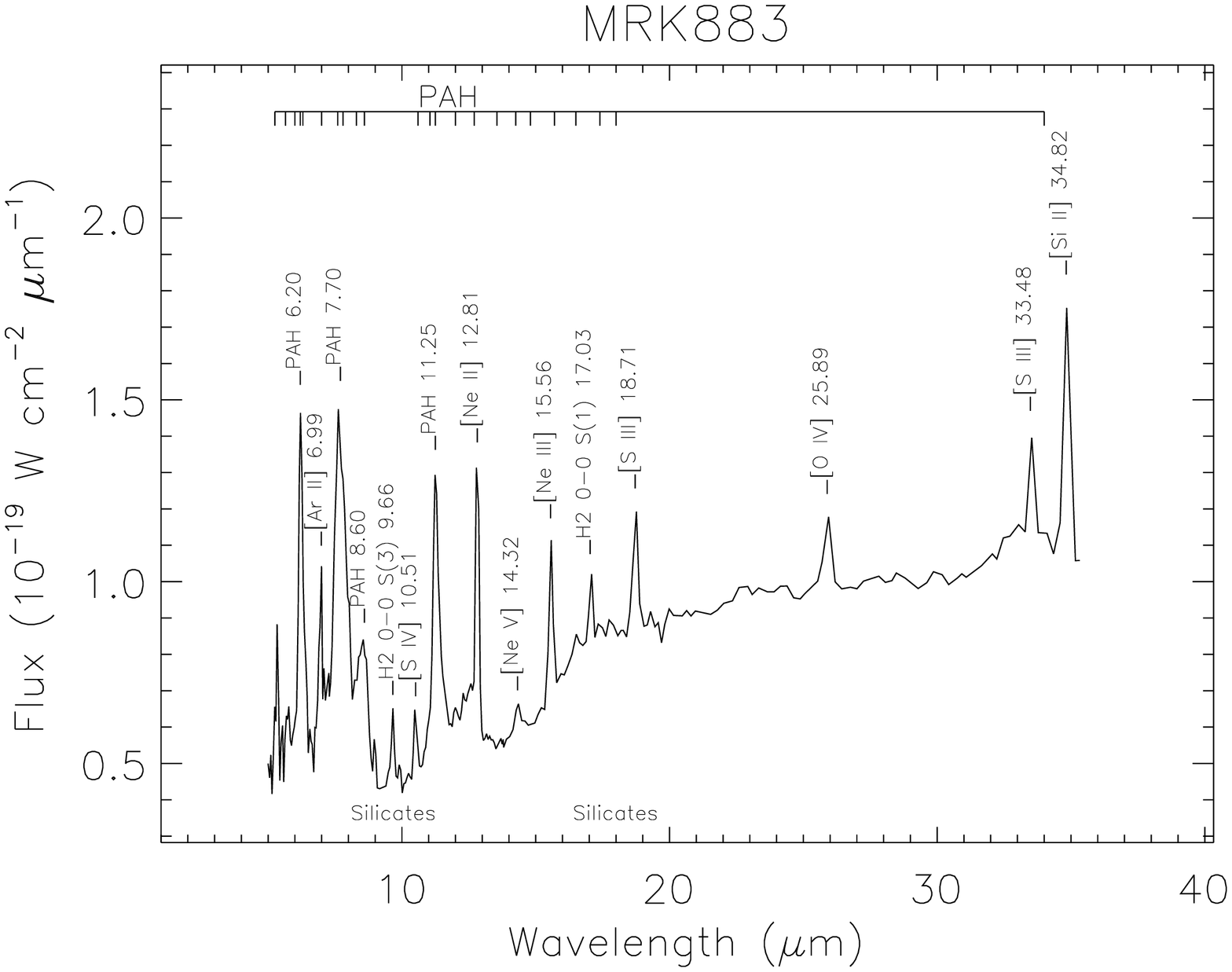}\\
    \includegraphics[width=0.45\textwidth,angle=90]{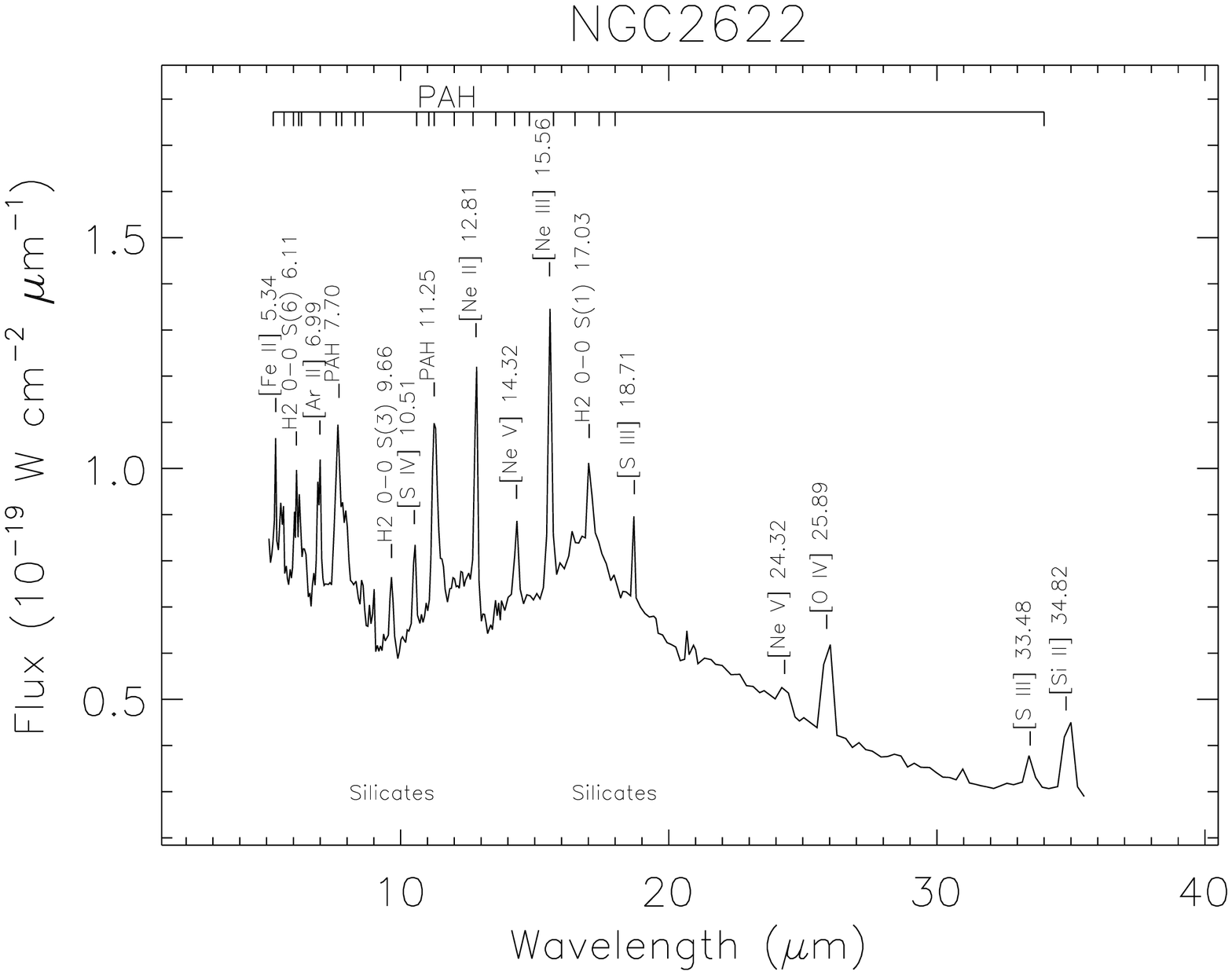}\\
    \caption{Continued -- Spectra of Seyferts from our new observations}
  \end{center}
\end{figure}
\clearpage

\setcounter{figure}{0}
\begin{figure}[!ht]
  \vspace{-2.0cm}
  \begin{center}
    \includegraphics[width=0.45\textwidth,angle=90]{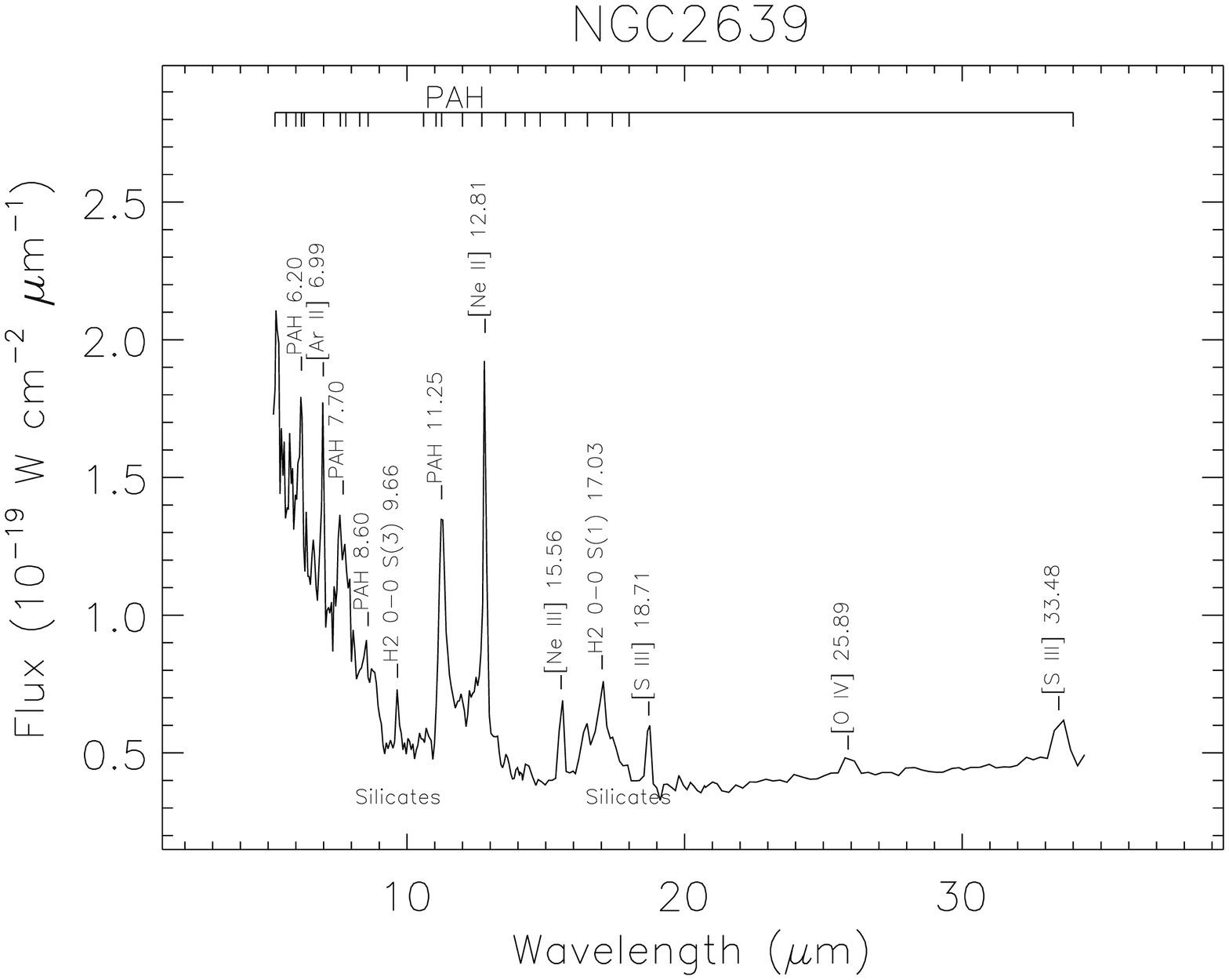}\\
    \includegraphics[width=0.45\textwidth,angle=90]{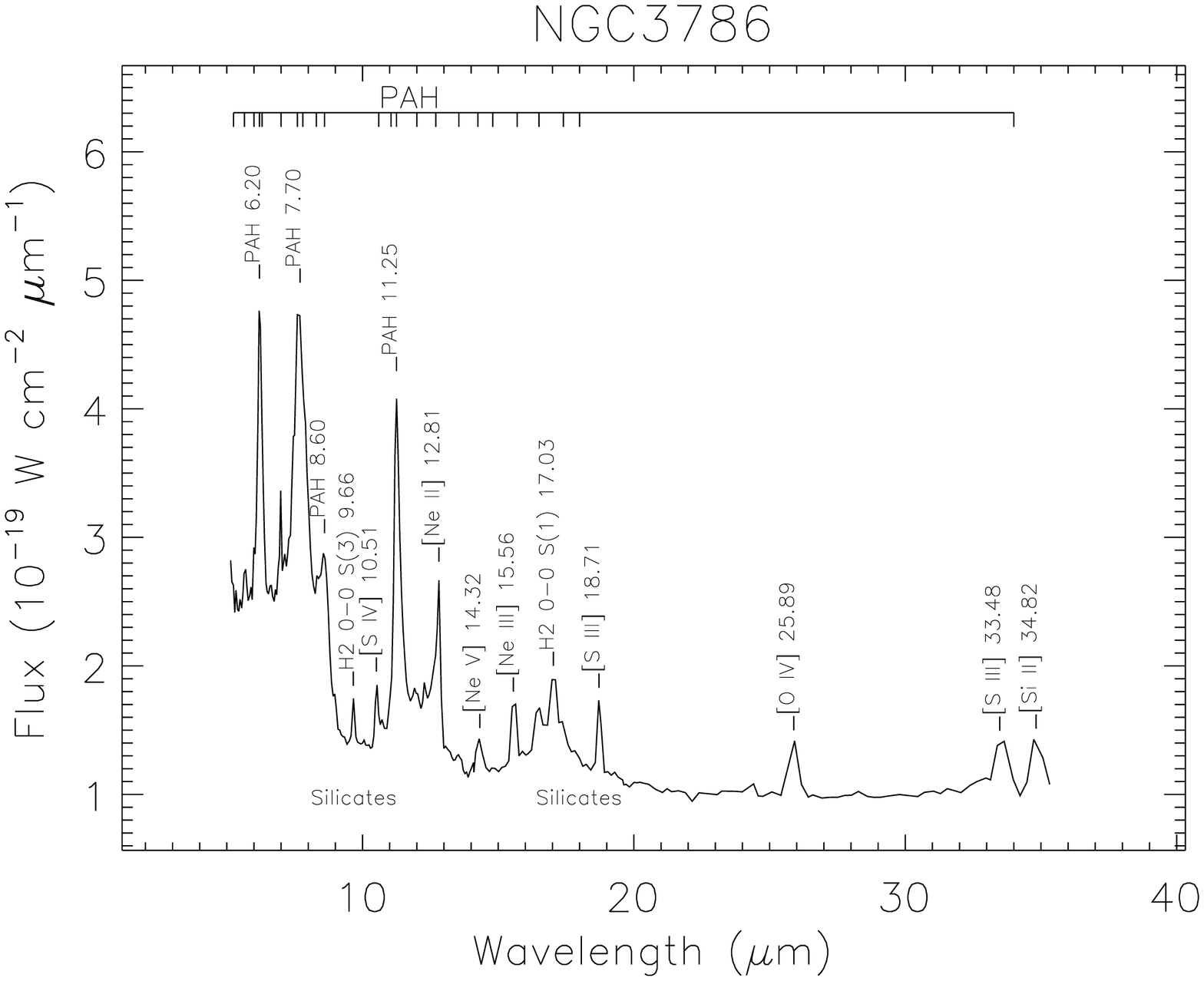}\\
    \includegraphics[width=0.45\textwidth,angle=90]{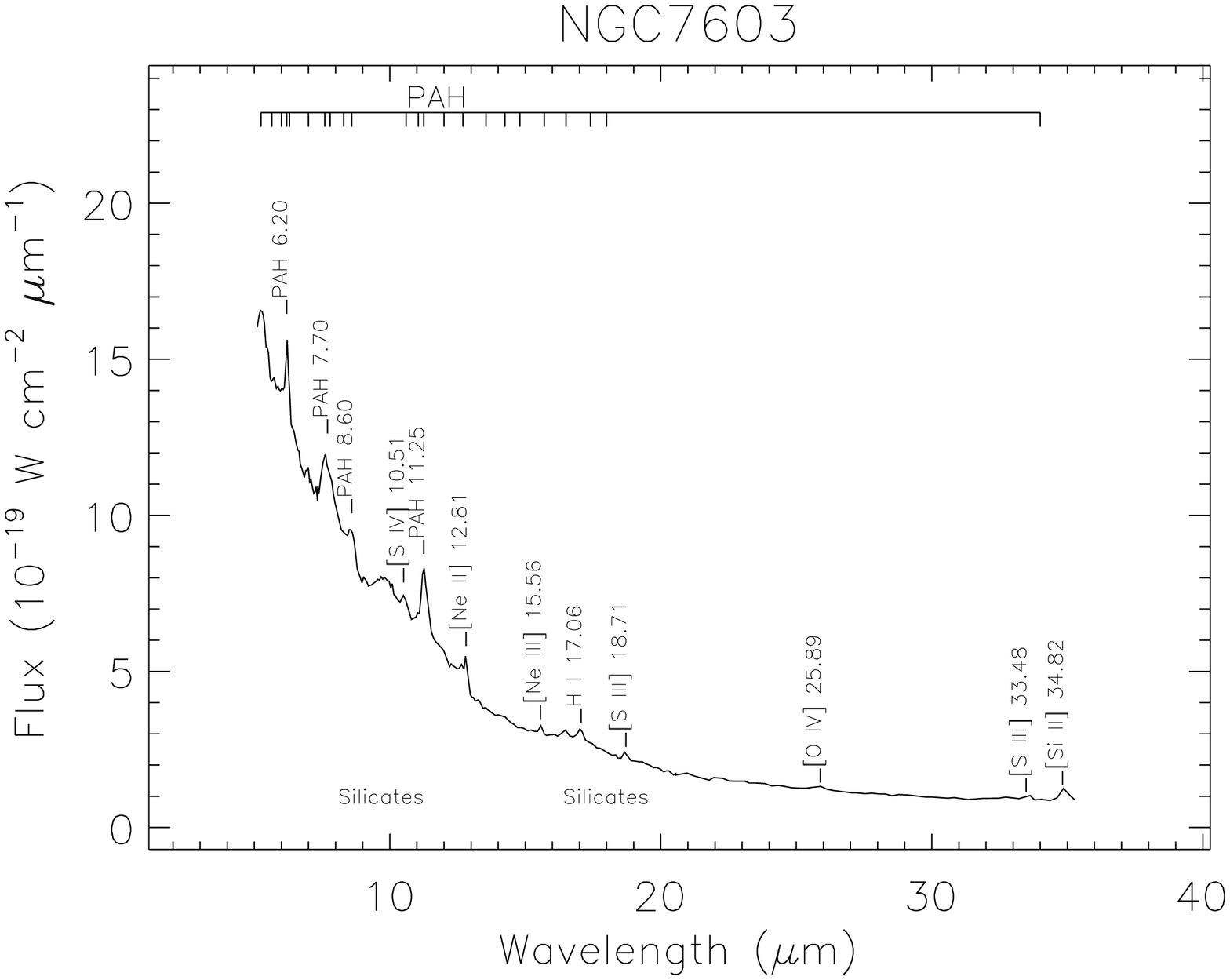}\\
    \caption{Continued -- Spectra of Seyferts from our new observations}
  \end{center}
\end{figure}
\clearpage

\setcounter{figure}{0}
\begin{figure}[!ht]
  \vspace{-2.0cm}
  \begin{center}
    \includegraphics[width=0.45\textwidth,angle=90]{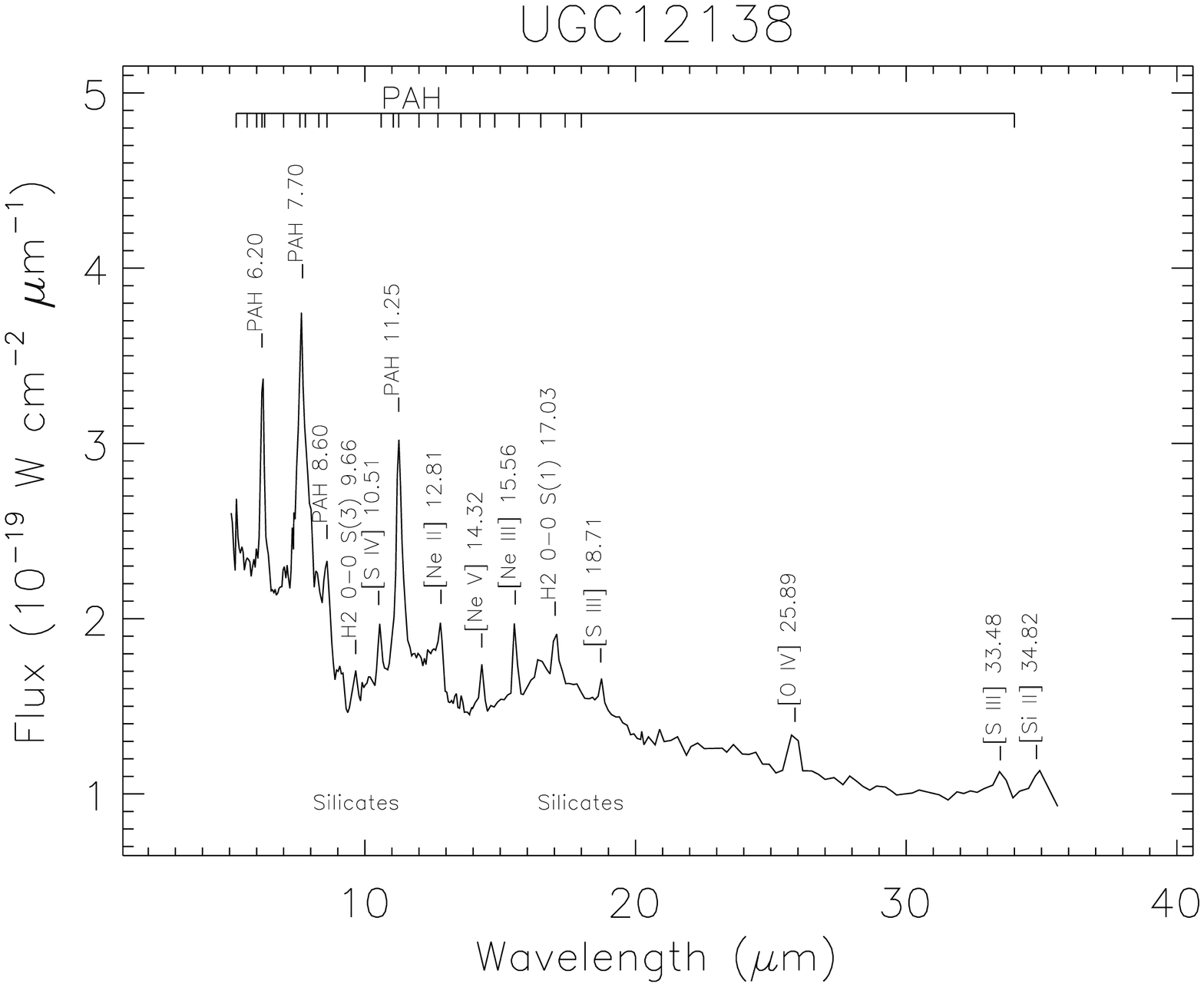}\\
    \includegraphics[width=0.45\textwidth,angle=90]{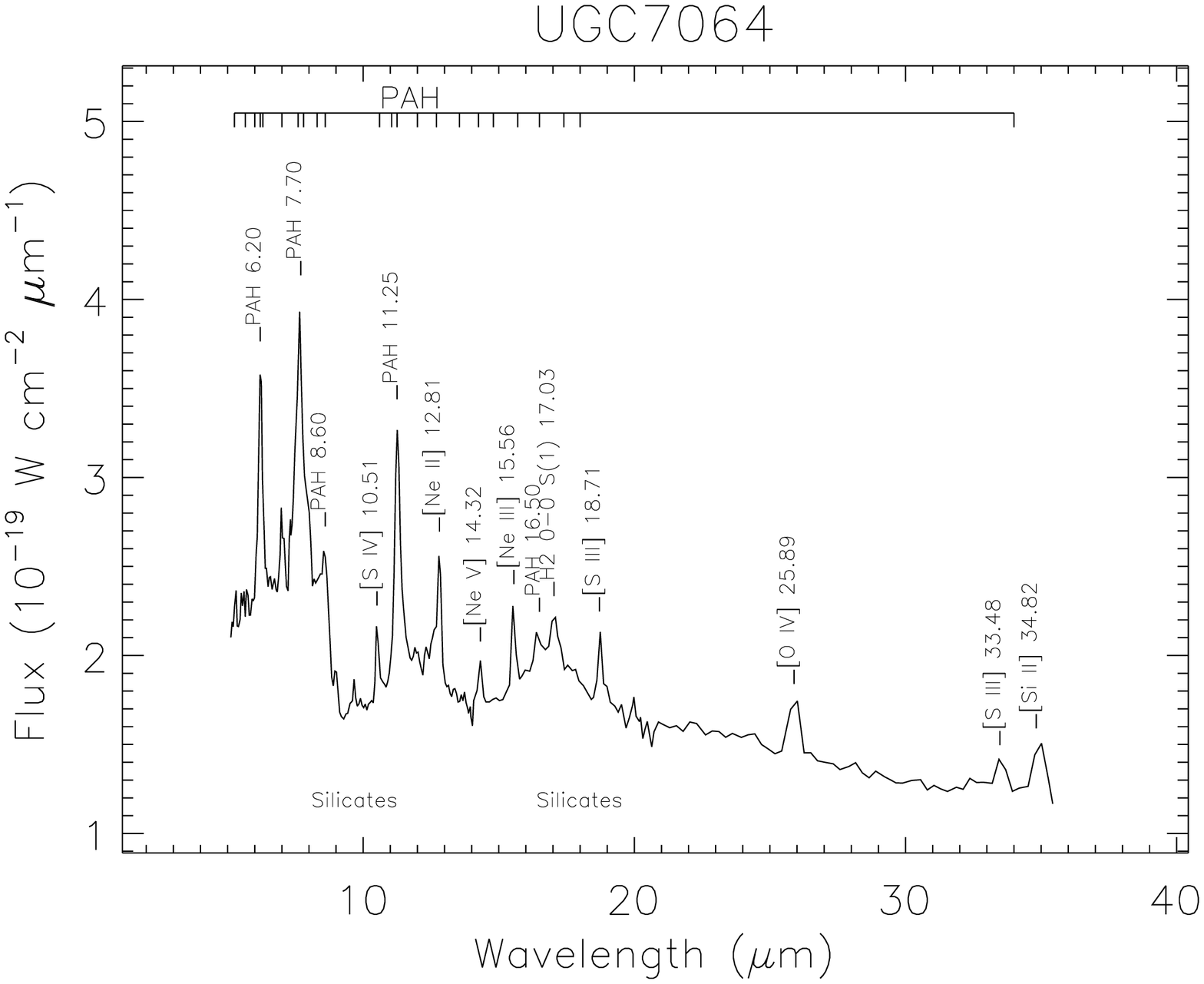}\\
    \includegraphics[width=0.45\textwidth,angle=90]{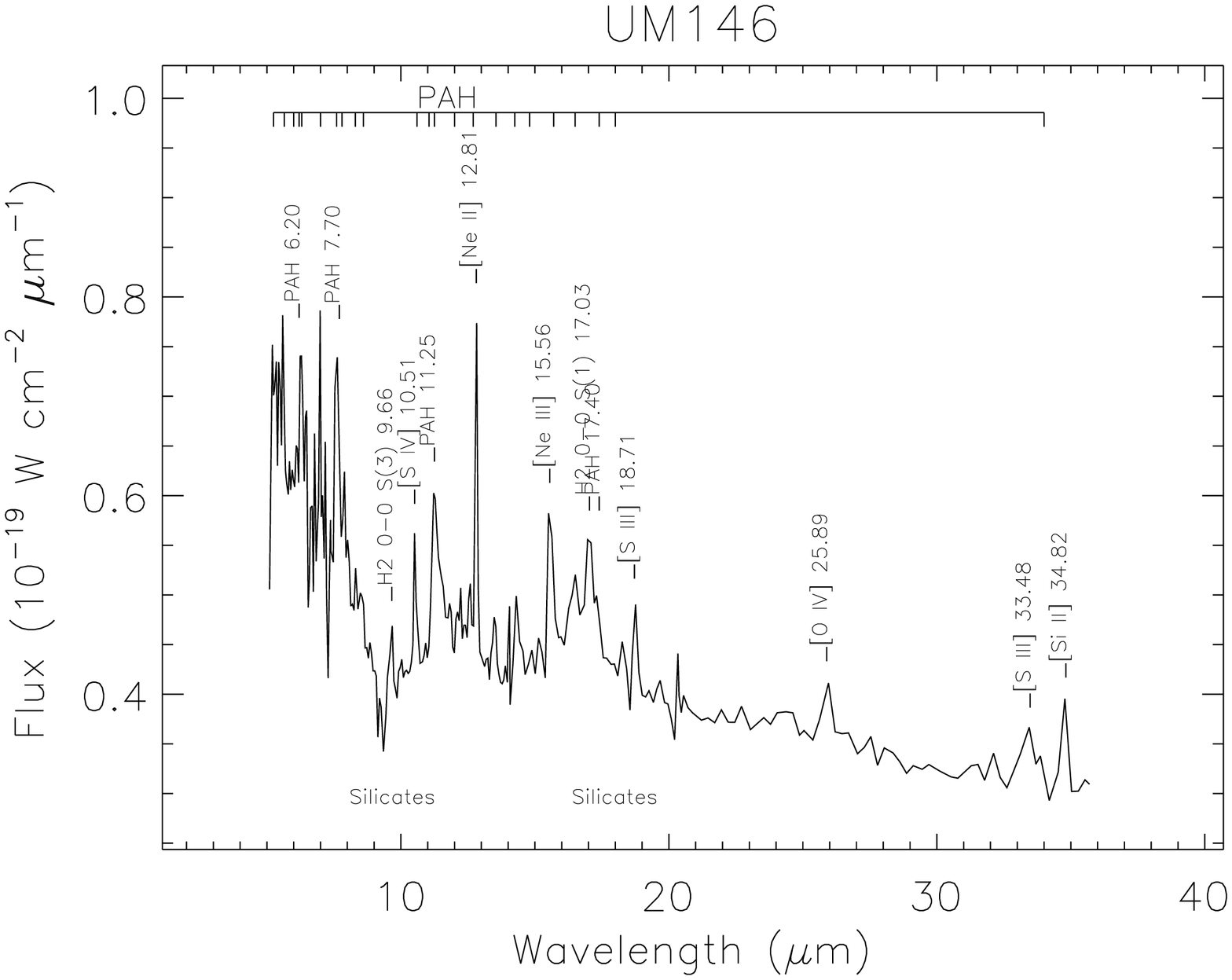}\\
    \caption{Continued -- Spectra of Seyferts from our new observations}
  \end{center}
\end{figure}
\clearpage

% Figure 2:  Power-laws
% \setcounter{figure}{0}
\begin{figure}[!ht]
  \begin{center}
    \includegraphics[angle=90,width=5.8in,keepaspectratio]{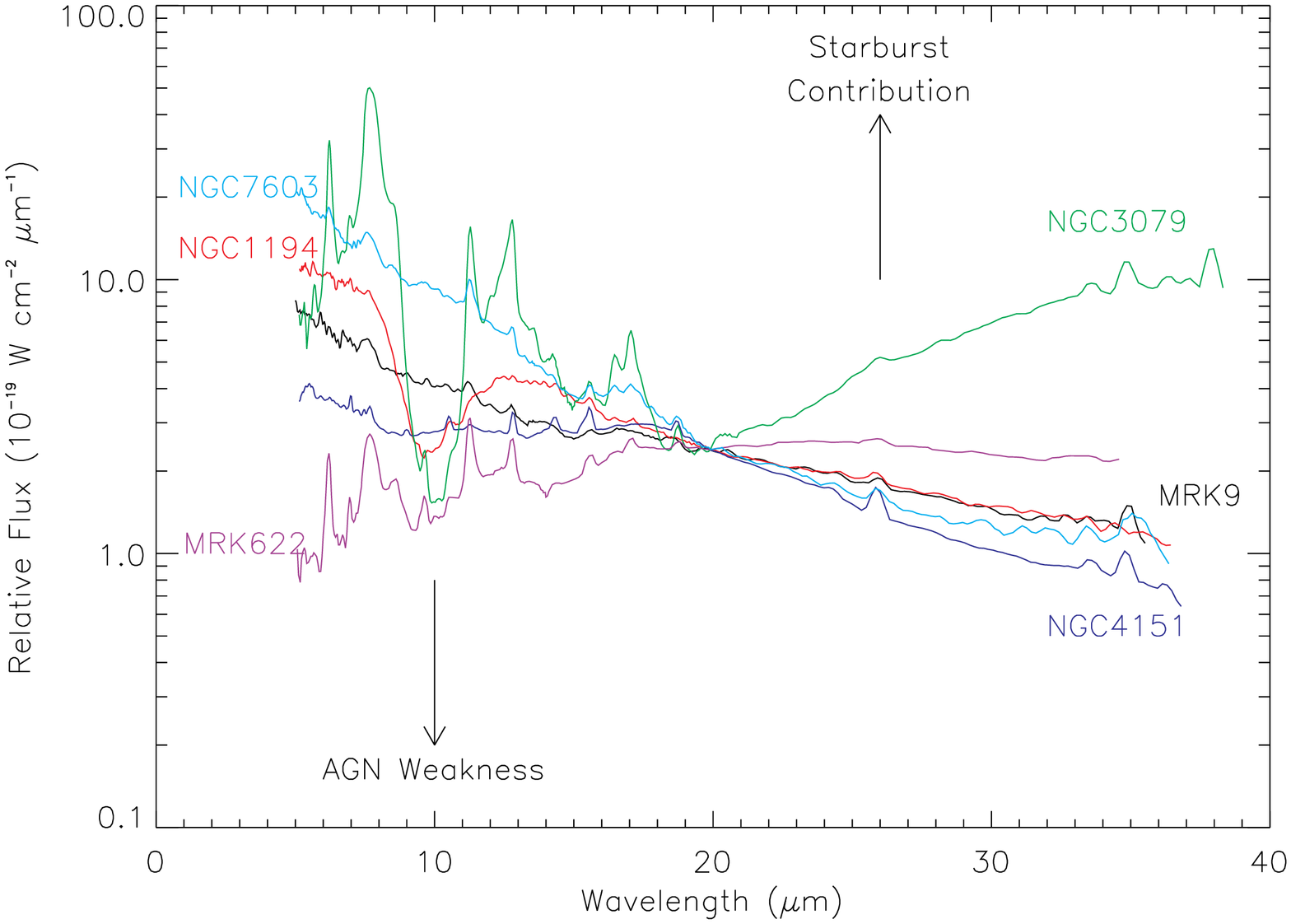}
    \caption{Variations in AGN continuum:  Mrk 9 (Seyfert 1.5, single power
      law --- SP), NGC 1194 (Seyfert 1, single power law with silicate
      absorption), NGC 4151 (Seyfert 1.5, broken power law --- BP), Mrk 622
      (Seyfert 2, broken power law --- BP), NGC 3079 (Seyfert 2, strong PAH,
      strong red continuum --- RC), and NGC 7603 (Seyfert 1, unusual quasar
      like spectrum, strong silicate emission feature at $10\mum$). The
      spectra have been normalized to the flux at $20\mum$ and smoothed by a
      factor of 2. The right arrow indicates that as the $20\textrm{--}30\mum$
      spectral index becomes positive, the mid-IR spectrum is more and more
      dominated by starburst features like PAH bands. The left arrow shows the
      amount of variations in the $6\textrm{--}15\mum$ spectral index, as we
      go down from NGC 7603 to Mrk 622 the contribution of hottest dust
      decreases. Optical Seyfert classifications do not always agree with the
      shape of the mid-IR spectrum.}
  \end{center}
\end{figure}
\clearpage

% Figure 3
\begin{figure}[!ht]
  \begin{center}
    \includegraphics[angle=90,width=\textwidth,keepaspectratio]{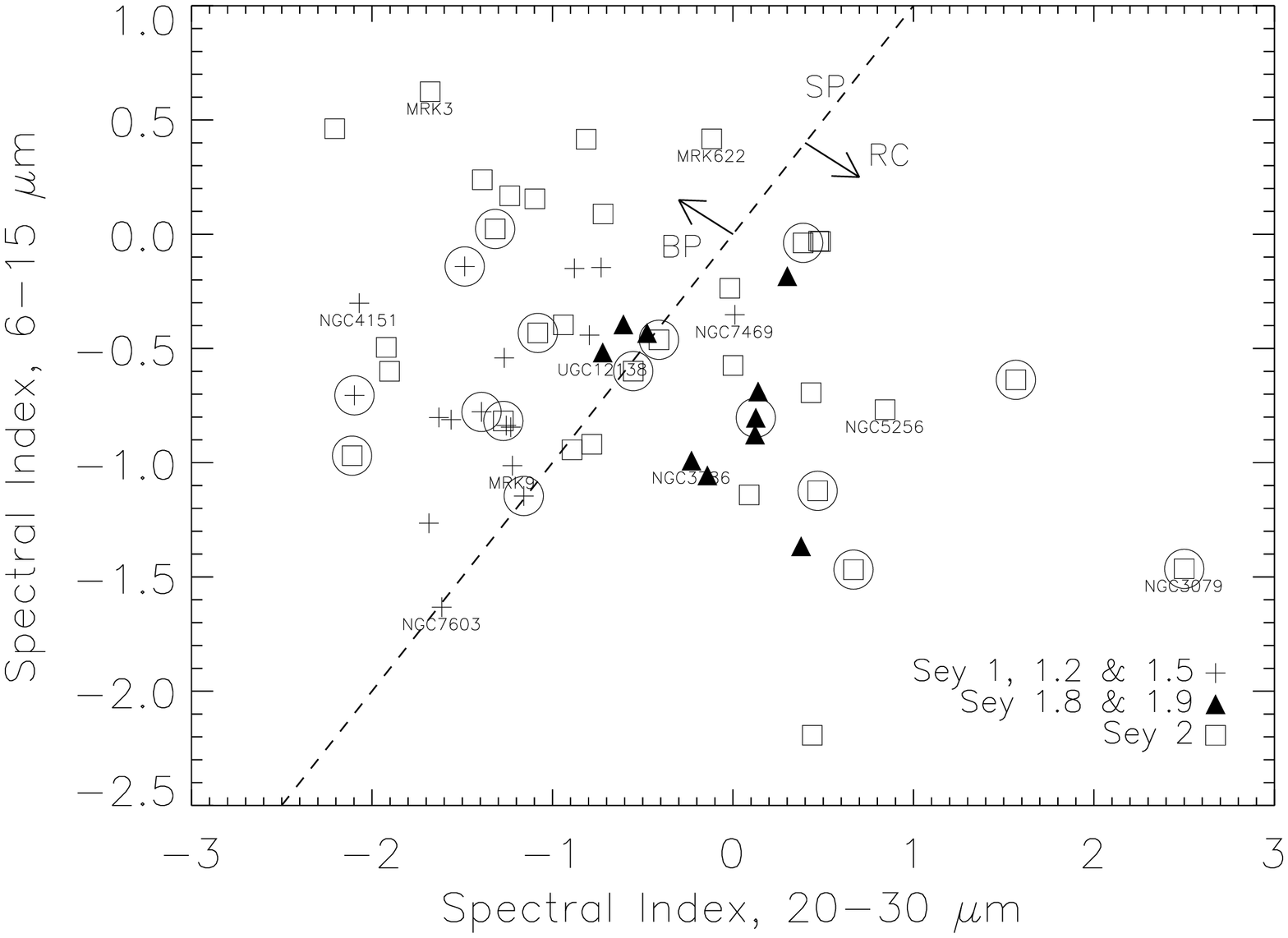}
    \caption{The $6\textrm{--}15\mum$ spectral index \vs $20\textrm{--}30
      \mum$ spectral index. BP --- Broken Power law, SP --- Single Power law
      and RC --- Red Continuum. The circled symbols have host galaxies with
      $b/a \le 0.5$ and/or are interacting. See Figure~2 for example spectra.}
  \end{center}
\end{figure}
\clearpage

% Figure 4
\begin{figure}[!ht]
  \begin{center}
    \includegraphics[angle=90,width=\textwidth,keepaspectratio]{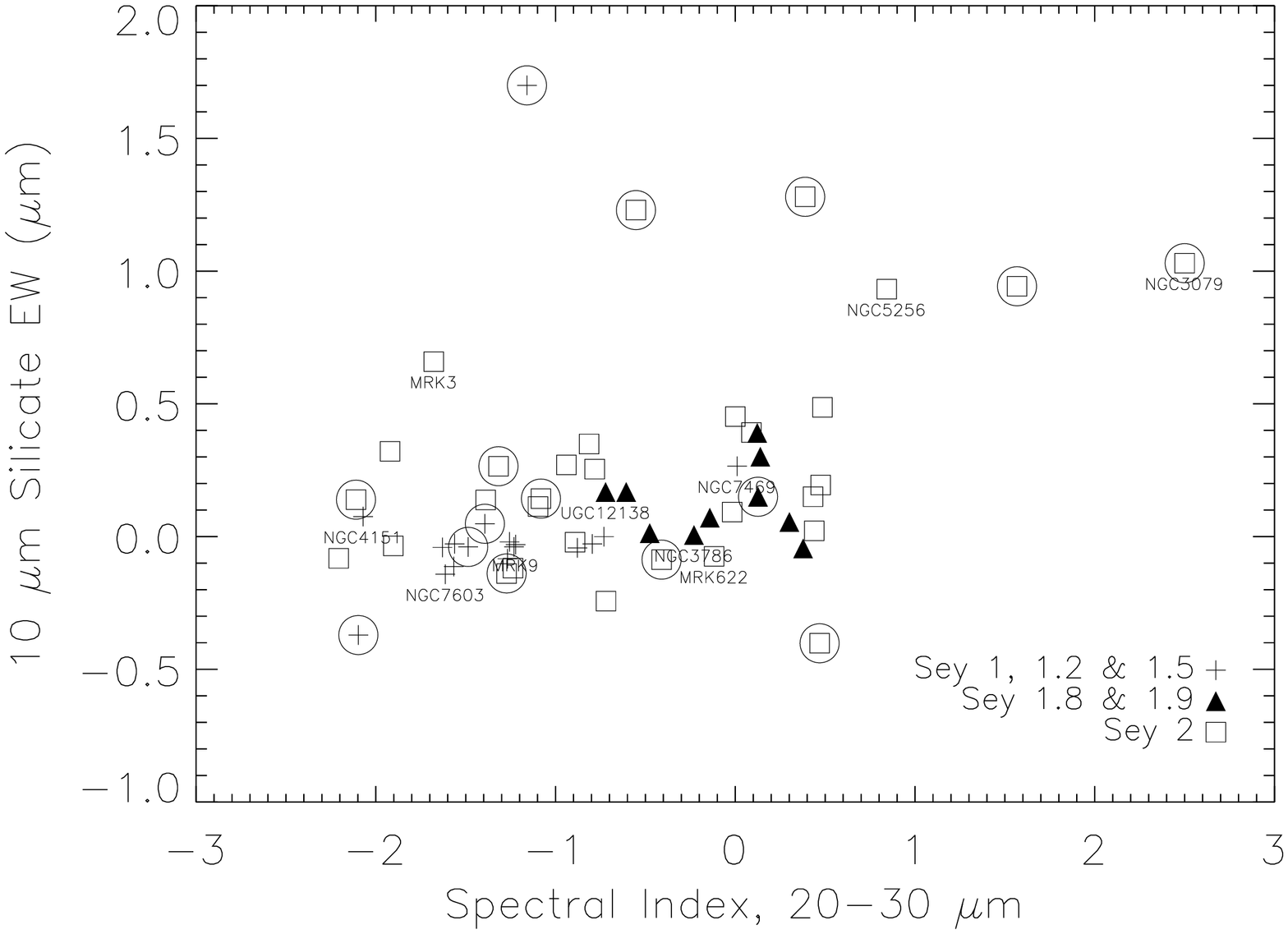}
    \caption{Silicate equivalent width ($10\mum$) \vs $20\textrm{--}30\mum$
      spectral index. The circled symbols have host galaxies with $b/a \le
      0.5$ and/or are interacting.}
  \end{center}
\end{figure}
\clearpage

% Figure 5
\begin{figure}[!ht]
  \begin{center}
    \includegraphics[angle=90,width=\textwidth,keepaspectratio]{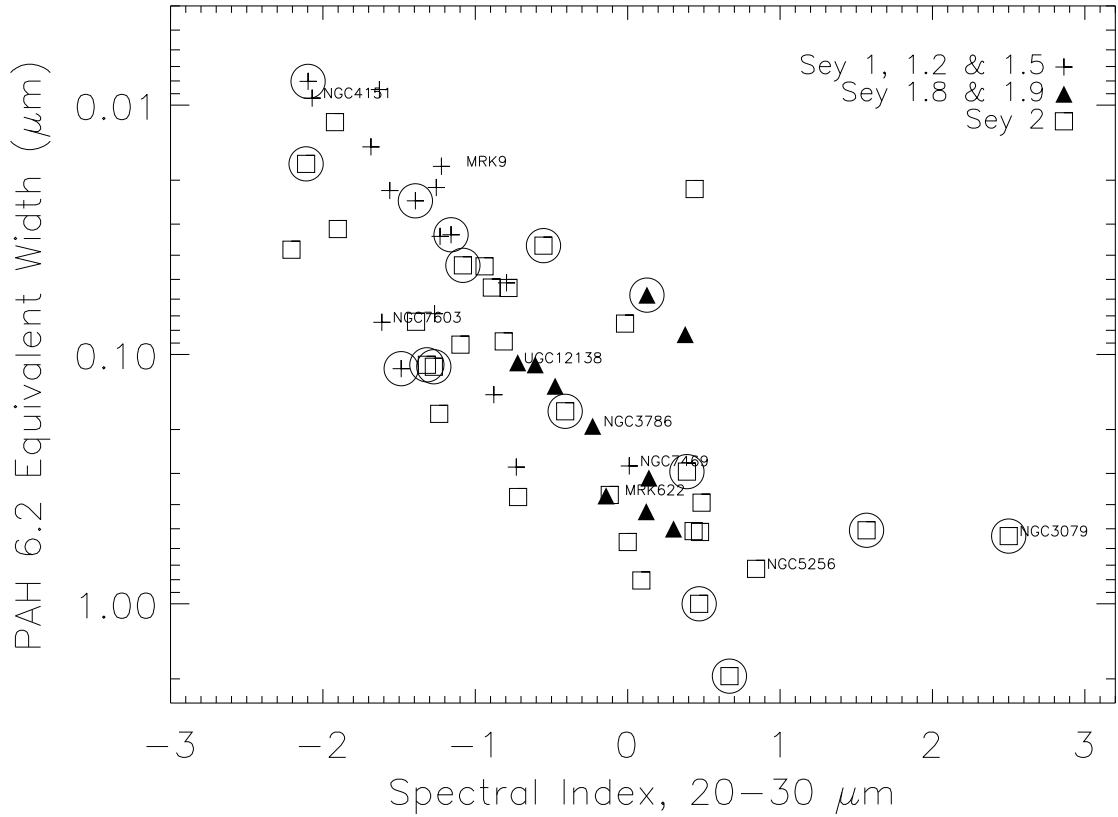}
    \caption{The $6.2\mum$ PAH equivalent width \vs the $20\textrm{--}30 \mum$
      spectral index. Equivalent widths are negative for emission features and
      absolute values are plotted here in log. Going from left to right on the
      x-axis, starburst contribution to the spectrum increases. The circled
      symbols have host galaxies with $b/a \le 0.5$ and/or are interacting.}
  \end{center}
\end{figure}
\clearpage

% Figure 6
\begin{figure}[!ht]
  \begin{center}
    \includegraphics[angle=90,width=\textwidth,keepaspectratio]{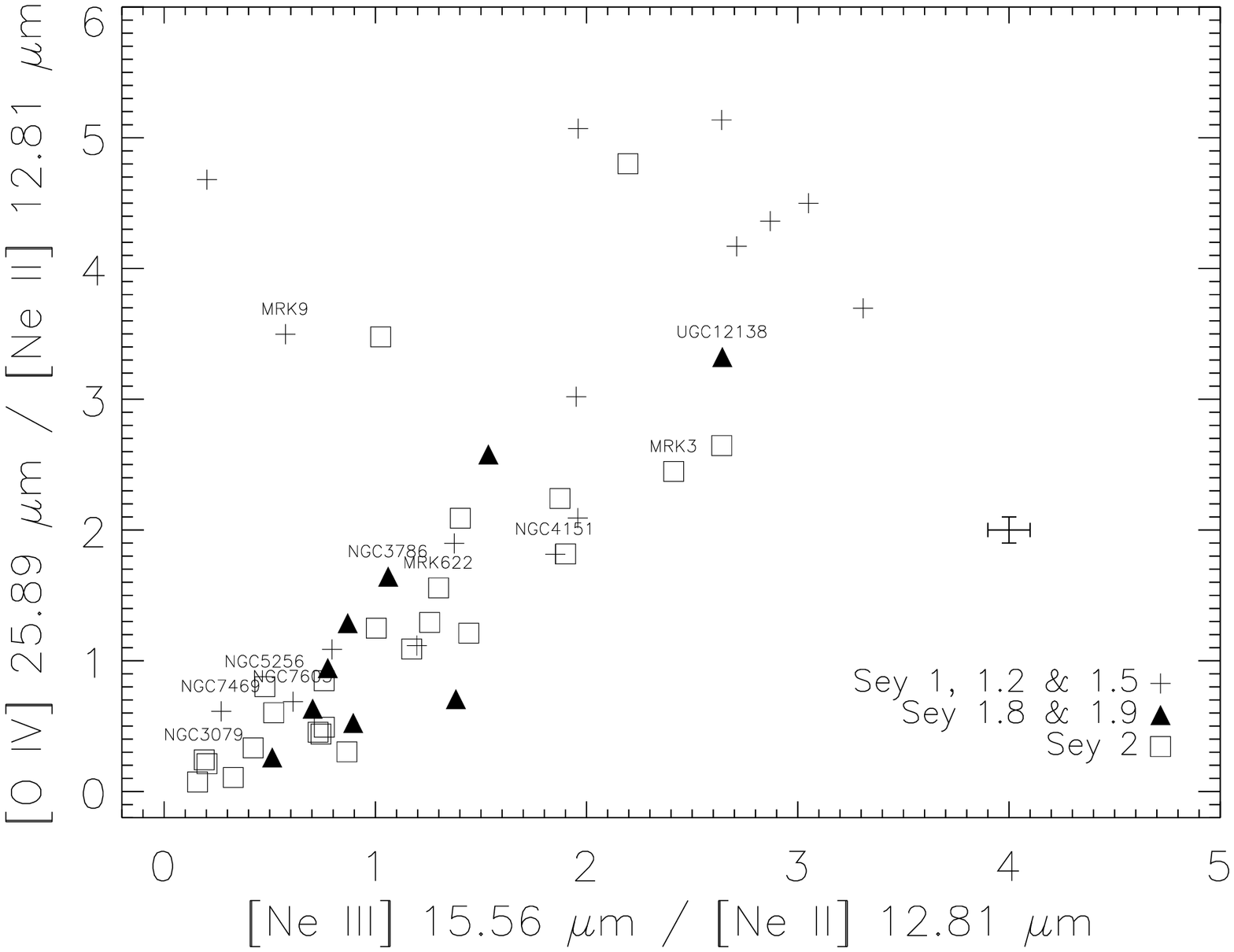}
    \caption{The [O IV] $25.89$ /[Ne II] $12.81$ ratio \vs [Ne III] $15.56$
      /[Ne II] $12.81$ ratio. Correlation between Seyfert 1.8-1.9s and Seyfert
      2s indicates that both subclasses of Seyfert galaxies in this sample
      have similar amounts of NLR ionization or they have intrinsically
      stronger starburst contribution as compared to Seyfert 1s.}
  \end{center}
\end{figure}
% \clearpage  

\end{document}